\documentclass[a4paper,fleqn]{cas-dc}
\usepackage{booktabs}
\usepackage{amsmath}
\DeclareMathOperator*{\argmin}{arg\,min}
\usepackage{siunitx}
\usepackage{upgreek}
\usepackage{wrapfig}

\usepackage{algorithm}
\usepackage{algpseudocode}

\usepackage[accsupp]{axessibility}  
\usepackage{bm}
\usepackage{multirow}
\usepackage{tikz}
\usepackage{siunitx}
\usepackage{comment}
\usepackage{amsmath,amssymb} 
\usepackage{color}
\usepackage{pifont}
\usepackage{makecell}
\usepackage{caption}
\usepackage{float}
\usepackage{adjustbox}
\usepackage{etoolbox}
\usepackage{dsfont}

\usepackage[accsupp]{axessibility}  
\usepackage{paralist}

\definecolor{rblue}{rgb}{0,0.5,1}

\usepackage{amsmath,lipsum}
\usepackage{cuted}
\usepackage{lineno}
\usepackage{booktabs,makecell,multirow}
\usepackage{amsmath,amssymb,amsfonts}
\usepackage{graphicx}
\usepackage{textcomp}
\usepackage{amsmath,amsfonts}
\usepackage{array}
\usepackage[caption=false,font=normalsize,labelfont=sf,textfont=sf]{subfig}
\usepackage{textcomp}
\usepackage{stfloats}
\usepackage{url}
\usepackage{verbatim}
\usepackage{graphicx}
\usepackage{cite}
\usepackage{booktabs}
\usepackage{paralist}

\usepackage{threeparttable}

\usepackage[numbers,sort&compress]{natbib}

\def\tsc#1{\csdef{#1}{\textsc{\lowercase{#1}}\xspace}}
\tsc{WGM}
\tsc{QE}

\usepackage{xspace}
\makeatletter
\DeclareRobustCommand\onedot{\futurelet\@let@token\@onedot}
\def\@onedot{\ifx\@let@token.\else.\null\fi\xspace}

\makeatother

\usepackage{color}

\usepackage{xcolor}
\definecolor{hollywoodcerise}{rgb}{0.96, 0.0, 0.63}
\definecolor{lasallegreen}{rgb}{0.03, 0.47, 0.19}
\definecolor{hanpurple}{rgb}{0.32, 0.09, 0.98}
\definecolor{green(pigment)}{rgb}{0.0, 0.65, 0.31}

\hypersetup{colorlinks=true,linkcolor={red},citecolor={hanpurple},urlcolor={magenta}} 

\usepackage{caption}
\begin{document}
\let\WriteBookmarks\relax
\def\floatpagepagefraction{1}
\def\textpagefraction{.001}

\shorttitle{Minimalist Biomimetic Computational Panoramic Imaging}    

\shortauthors{S. Gao \textit{et al.}}  

\title [mode = title]{Minimalist Biomimetic Computational Panoramic Imaging}  

\author[1]{Shaohua Gao}
\cormark[1]
\credit{Conceptualization of this study, Methodology, Software}
\author[1]{Qi Jiang}
\cormark[1]
\author[1]{Ning Pan}
\author[1]{Yao Gao}
\author[1]{Zhonghua Yi}
\author[1]{Ze Wang}
\author[1]{Xiaolong Qian}
\author[1]{Biaofeng Zeng}
\author[2]{Kailun Yang}
\author[1]{Kaiwei Wang}[orcid=0000-0002-8272-3119]
\cormark[2]
\ead{wangkaiwei@zju.edu.cn}
\author[1]{Jian Bai}

\affiliation[a]{organization={State Key Laboratory of Extreme Photonics and Instrumentation, College of Optical Science and Engineering},
            addressline={Zhejiang University}, 
            city={Hangzhou},
            postcode={310027}, 
            country={China}}

\affiliation[b]{organization={National Engineering Research Center of Robot Visual Perception and Control Technology},
            addressline={Hunan University}, 
            city={Changsha},
            postcode={410082}, 
            country={China}}

\cortext[1]{Equal contribution}
\cortext[2]{Corresponding author}

\begin{abstract}
Conventional large field of view (FoV) panoramic imaging systems consist of multiple lenses to correct aberrations, making them bulky, heavy, and difficult to fabricate for use in intelligent sensing scenarios that are tightly constrained in size and weight, such as miniaturized drones and micro-robots. We are inspired by a four-eyed fish (Anableps anableps) in nature that can observe the FoVs above and below the water surface simultaneously. Here, we construct an ultra-compact biomimetic computational panoramic imaging system that consists only of a monolithic lens, an optical micro-channel glass array, and an image sensor. In addition, the imaging system includes a panoramic computational imaging algorithm to enhance image quality and improve its perception in the panoramic FoV. This study achieved ultra hemispherical panoramic imaging using only one lens optical architecture, achieving high-quality panoramic imaging with an image resolution of 3K on a 1-inch large image area. In terms of the results of the study, this system has the advantages of ultra-large FoV, small size, light weight, minimalist architecture, simple manufacturing, and low cost, which is expected to bring promising application prospects in machine vision perception scenarios with strict size and weight constraints.
\end{abstract}

\begin{keywords}
 \sep Optical Design
 \sep Minimalist Panoramic Imaging
 \sep Optical Information Processing
 \sep Computational Imaging
 \sep Computational Aberration Correction
\end{keywords}
\maketitle

\section{Introduction}
Various lens designs have flourished since the invention of the Daguerreotype's photographic lens in 1839~\cite{kingslake1989history}. As the core components of the imaging system, the lens surface types have ranged from spherical surface to aspheric surface~\cite{zhu2022monolithic} to freeform surface~\cite{rolland2021freeform}, and the light transmission methods have ranged from refraction to diffraction~\cite{banerji2019imaging}. 
Scientists have made massive research and great contributions to the new architecture, materials, functions, and performance of lenses~\cite{yu2014flat,khorasaninejad2017metalenses,bauer2018starting}. 
 
At present, traditional small field of view (FoV) bulky imaging lenses are no longer able to meet the fast movement perception capabilities of various miniaturized unmanned aerial vehicles and micro robots in the field of ultra-low power consumption requirements. The characteristics of miniaturized~\cite{dallaire2015wide,cheng2021freeform,pan2022dielectric,miller2023optics}, lightweight~\cite{dorrah2022tunable,scherrer2023optical}, high-quality~\cite{yang2017automated,liang2019high}, and large FoV~\cite{chen2022planar,xie2023ultrathin} lenses have become an urgent demand in the field of imaging perception for micro intelligent mobile terminals.
 
The panoramic imaging system, as an optical device that can achieve $360^\circ$ hyper-hemispheric FoV imaging, is widely used in many fields such as video conferencing, scene monitoring, environmental perception,  augmented reality (AR) / virtual reality (VR) panoramic video acquisition, and machine vision~\cite{thibault2008panoramic,gao2022review}. However, due to the large Petzval field curvature in the optical aberration brought about by large FoV imaging, panoramic imaging systems usually require 7$\sim$11 lenses to correct off-axis aberration to achieve better imaging quality~\cite{wang2019design,zhang2020design,zhou2020design,pernechele2021telecentric,wang2022high}. Therefore, panoramic imaging systems inevitably still suffer from the problems of higher lens counts, system complexity, large size, and heavy weight, and cannot be applied in bulk to perception scenarios such as miniaturized unmanned aerial vehicles (UAVs), miniaturized robots, and other space- and weight-strictly constrained sensing scenarios.

In recent years, biomimetic optics have inspired the development of new compact large FoV imaging optical systems~\cite{lee2005inspirations}. 
To achieve a compact and minimalist optical system, biomimetic human eyes~\cite{lee2018bioinspired}, aquatic vision~\cite{kim2020aquatic}, and insect compound eyes~\cite{liang2019one,zhou2024ultrawide} use concentric optical systems~\cite{kim2020aquatic}, fiber bundle imaging~\cite{stamenov2014panoramic}, curved surface sensors~\cite{zhang2017origami}, and array sensors~\cite{zhou2024ultrawide} as new optical component imaging devices, greatly simplifying the imaging system architecture of traditional panoramic systems that use planar image sensors as information receiving devices~\cite{wang2019design}. The origami silicon optoelectronics imaging~\cite{zhang2017origami} and multi-image sensor stitching technology simplify the complexity and design difficulty of subsystems. Although the optical architecture of the imaging system has been simplified, precise and complex low-resolution novel sensors are still needed for imaging perception. The design of the array requires pre-calibration and calibration of each subsystem, and sub-image stitching loses the multiplexed imaging FoVs, and stitching traces are not conducive to the panoramic perception of large FoVs. 
In nature, a ``four eyed fish'' (Anableps anableps) has evolved an eyeball conjunctiva to simultaneously observe predators and prey on and below the water surface. The conjunctiva divides a single fish eye into upper and lower parts. Due to the protruding structure of the eyeball, the observation area of the unilateral FoV of the ``four eyed fish'' is the near hemisphere FoV without significantly increasing the volume of the eyeball. It can simultaneously observe a large FoV above and below the water surface, providing a promising idea for the development of new panoramic imaging system architectures. Based on its imaging principle, this paper constructs an ultra-compact biomimetic computational panoramic imaging system, which consists only of a single new type of catadioptric lens, a micro-channel array, and an image sensor. It also includes a panoramic computational imaging algorithm to enhance image quality and improve its perception ability. The system has the advantages of a hyper-hemispherical FoV, small size, light weight, minimalist architecture, simple manufacturing, and low cost, and is expected to be applied in machine vision perception scenarios with strict limitations on volume and weight. 
In Table~\ref{tab:1_}, we summarize the most advanced representative compact large-FoV biomimetic imaging systems of the past decade.

\begin{table*}[!ht]
\centering
\scriptsize
\begin{threeparttable}
\caption{Comparison of state-of-the-art compact large FoV biomimetic imaging systems.\tnote{\textit{a}}}
\setlength{\tabcolsep}{0.5mm}{
\begin{tabular}{cccccc}
\toprule
\makecell[c]{System
}&\makecell[c] {HFoV
($^{\circ}$)}&\makecell[c]{F/\#
}&\makecell[c]{Lenses number}&\makecell[c]{Imaging method}&\makecell[c]{Image imaging mode}\\
\midrule
\makecell[c]{Opt. Express (2014)~\cite{stamenov2014panoramic}}&\makecell[c]{0$\sim $8/63}&\makecell[c]{1.0}&\makecell[c]{5}&\makecell[c]{Curved fiber-coupled sensors
}&\makecell[c]{Sub-image stitching}
\\
\midrule
\makecell[c]{Nat. Communications (2017)~\cite{zhang2017origami}}&\makecell[c]{0$\sim $15}&\makecell[c]{1.5}&\makecell[c]{1}&\makecell[c]{Curved sensor
}&\makecell[c]{Silicon optoelectronics}
\\
\midrule
\makecell[c]{Optica (2019)~\cite{liang2019one}}&\makecell[c]{0$\sim $46}&\makecell[c]{1.7}&\makecell[c]{1}&\makecell[c]{Curved sensor
}&\makecell[c]{Sub-image stitching}
\\
\midrule
\makecell[c]{SPIE (2019)~\cite{wang2019design}}&\makecell[c]{30$\sim $100}&\makecell[c]{4.0}&\makecell[c]{9}&\makecell[c]{Flat sensor
}&\makecell[c]{Direct imaging}
\\
\midrule
\makecell[c]{Nat. Electronics (2020)~\cite{kim2020aquatic}}&\makecell[c]{0$\sim $60}&\makecell[c]{2.6}&\makecell[c]{4}&\makecell[c]{Curved sensor
}&\makecell[c]{Direct imaging}
\\
\midrule
\makecell[c]{Science Robotics (2024)~\cite{zhou2024ultrawide}}&\makecell[c]{0$\sim $70/110}&\makecell[c]{Pinhole}&\makecell[c]{121}&\makecell[c]{Array sensors
}&\makecell[c]{Sub-image stitching}
\\\midrule
\makecell[c]{This Work A}&\makecell[c]{40$\sim $95}&\makecell[c]{5.3}&\makecell[c]{1}&\makecell[c]{Flat micro-channel sensor
}&\makecell[c]{Direct imaging}
\\\midrule
\makecell[c]{This Work B}&\makecell[c]{40$\sim $100}&\makecell[c]{2.8}&\makecell[c]{1}&\makecell[c]{Flat micro-channel sensor
}&\makecell[c]{Direct imaging}
\\

\bottomrule
\label{tab:1_}
\end{tabular}}

\begin{tablenotes}
        \footnotesize
        \item[\textit{a}] /: It represents that the horizontal FoV of the system is not consistent with the vertical FoV. The half FoVs on the left and right sides of the symbol / are respectively the horizontal half FoV and the vertical half FoV.
      \end{tablenotes}
      \small
    \end{threeparttable}
\end{table*}

Compared with other state-of-the-art compact large FoV biomimetic imaging systems in the past decade, this work greatly reduces the number of lenses in the large FoV biomimetic imaging system, simplifies the imaging mode, and proposes a biomimetic panoramic system that combines the unique advantages of the four-eyed fish visual imaging optical path and high-sensitivity rod cell retina.

Research work A focuses on a minimalist biomimetic panoramic imaging system (MBPIS) with a $360^\circ{\times}(40^\circ{\sim}95^\circ$) FoV. To verify the correctness of the design concept, the F/\# in research work A was manufactured with a focal length of 3.7 mm, a single lens weight of 19.8 g, and a size comparable to a coin. The total optical length of the system is 14 mm, much smaller than traditional panoramic systems with the same target surface, achieving ultra-wide FoV imaging in a minimalist optical architecture and compact size. 
The MBPIS is equipped on a smartphone for imaging to verify that the optical system can also be mounted on small intelligent mobile computing terminals such as drones and miniaturized robots. The prototype's image sensor adopts a 1-inch ultra-large target area mobile phone image sensor, with a resolution of 8192 (H) $\times$ 6144 (V) and a single pixel size of 1.6 $\upmu$m.
In addition, this article also proposes the design of an aspherical large aperture MBPIS for research work B. Its F/\# can be as small as 2.8, the FoV reaches $360^\circ{\times}(40^\circ{\sim}100^\circ$), and the MTF is higher than 0.2 at 100 lp/mm, achieving high-definition panoramic images without the need for computational imaging. The minimalist architecture is achieved in terms of system complexity, realizing the design of a compact, high-definition, large-FoV panoramic imaging system.

Inspired by the regulatory function of highly sensitive rod cells in the retina of the four eyed fish under various strong and weak light conditions, this study achieved different F/\# imaging by using micro-channel arrays of different thicknesses, coupled with micro-channel arrays to form a new image sensor, and constructed a panoramic imaging perception constrained by the numerical aperture (NA) of the image square. In addition, the application of micro-channel arrays eliminates a large amount of stray light for this type of catadioptric system, achieving a unique physical performance of hardware-level stray light elimination, and solving the problem of strong stray light caused by large FoV and catadioptric architecture in traditional panoramic imaging systems. The minimalist biomimetic computational panoramic imaging framework developed in this study provides a new direction for the development of minimalist panoramic imaging systems, and brings broad application prospects for sensing environments with limited space and weight.

\section{Principles and Experimental Results}
\subsection{Principles of biomimetic design}
\begin{figure*}[!t]
  \centering
  \includegraphics[width=1.0\linewidth]{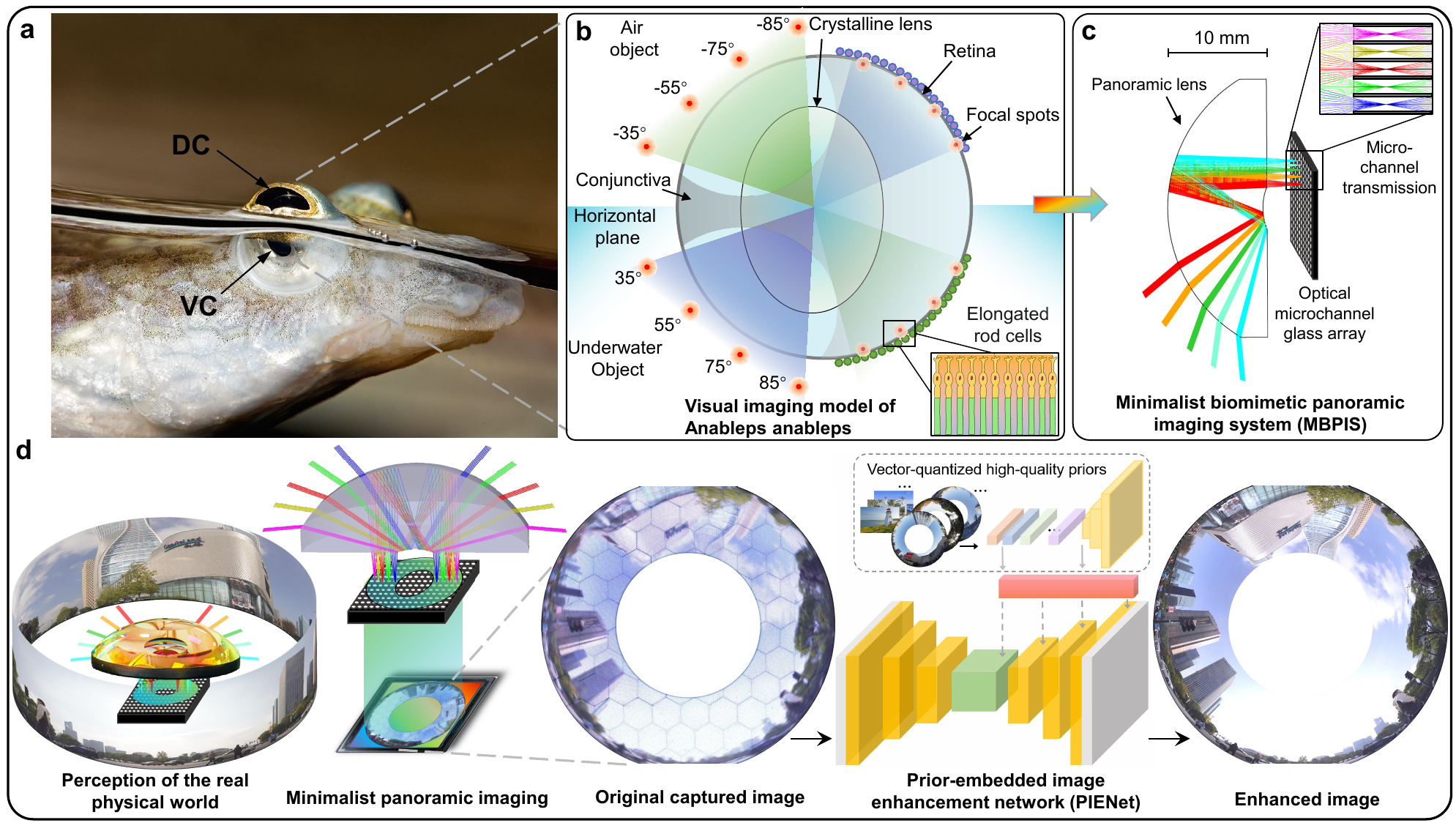}
  \caption{\textbf{The minimalist biomimetic computational panoramic imaging architecture and principle inspired by the four-eyed fish (Anableps anableps) Visual imaging model.} \textbf{a} The visual imaging model of the four-eyed fish (Anableps anableps). DC: Dorsal Cornea. VC: Ventral Cornea. Copyright obtained from Andreas Werth. \textbf{b} Visual imaging model of Anableps anableps. \textbf{c} A minimalist biomimetic panoramic imaging architecture consisting of a catadioptric single lens and an optical micro-channel glass array for ultra-large FoV imaging. \textbf{d} Minimalist biomimetic computational panoramic imaging principle.}
  \label{fig:Anableps anableps}
\end{figure*}

In Fig.~\ref{fig:Anableps anableps}a, a unique aquatic visual organism, the four-eyed fish, is shown on the water surface, with two eyes appearing on one side. In fact, each four-eyed fish individual only has two eyes, and each eye is divided into dorsal (upper) and ventral (lower) parts by a diaphragm tissue band~\cite{owens2012four,perez2017eye}. Fig.~\ref{fig:Anableps anableps}b shows a schematic cross-sectional view of four fisheye eyes, each with its own cornea and pupil, which can capture light from above and below the surface, while the lens can correct aberrations in different FoVs. The unusual optical anatomical structure evolved by the four-eyed fish can help it prey on both terrestrial and aquatic prey while avoiding attacks from predators. Due to long-term living in dim habitats, the retina of the four-eyed fish has evolved into highly sensitive rod cells separated into upper and lower parts, which can meet the high-definition perception of the water surface and underwater environment under low light conditions. The visual imaging structure and functional characteristics of the hyper-hemispherical large aperture in nature provide inspiration for the new generation panoramic imaging system shown in Fig.~\ref{fig:Anableps anableps}c. By transforming the diaphragm tissue band into a reflective mirror and using the optical path of secondary refraction and secondary reflection to compress the total length of the optical path, the axial length of the panoramic system can be greatly compressed. Coupling optical micro-channel glass arrays on sensors instead of high-sensitivity rod cells can achieve NA constraints on various FoVs, enabling large aperture imaging under different lighting conditions. This imaging architecture achieves a minimalist structure with a single lens, eliminating the need for the aperture stop commonly found in traditional panoramic systems. It enables the virtual aperture to stop imaging inside the lens, further reducing optical components and compressing the volume and weight of the system. By combining computational imaging technology, the perception ability of the system can be further enhanced. The framework for imaging perception of the minimalist biomimetic panoramic imaging system is shown in Fig.~\ref{fig:Anableps anableps}d. The 360{\textdegree} panoramic scene is imaged on the image plane by a single lens through a minimalist catadioptric architecture. The optical micro-channel glass array is coupled to the image sensor to constrain the image space NA and absorb stray light. Combining computational imaging techniques, we propose a prior-embedded image enhancement network (PIENet) to enhance images, thereby achieving high-quality panoramic imaging perception of the real physical world. The images obtained from the image sensor are used to calculate the imaging link for enhancement, ultimately obtaining high-definition panoramic images for scene perception.

\subsection{Design architecture comparison and prototype lens size comparison}
\label{sec:Design architecture comparison}

\begin{figure*}[!t]
  \centering
  \includegraphics[width=0.7\linewidth]{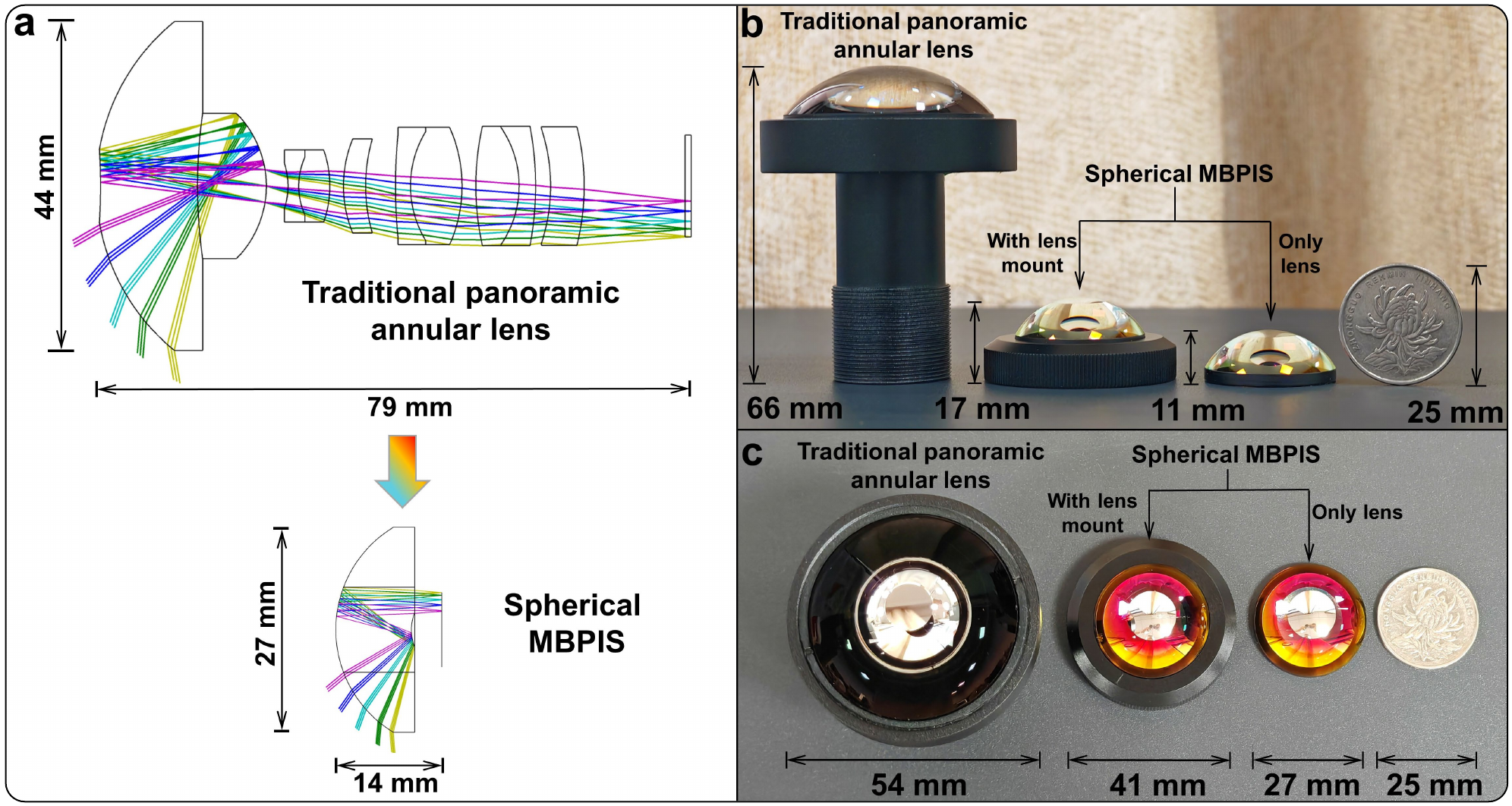}
  \caption{\textbf{Comparison between minimalist biomimetic computational panoramic imaging system and traditional panoramic annular lens.} \textbf{a} Optical design scheme. \textbf{b} Lens height. \textbf{c} Lens diameters.}
  \label{fig:Comparison}
\end{figure*}

The panoramic annular lens has been extensively studied due to its compact structure, small edge FoV distortion, and high relative illumination. However, it is still difficult to be widely used due to its large size, heavy volume, complex system structure, and inconvenient mass production. 
Compared with the traditional panoramic annular lens~\cite{jiang2024minimalist}, the spherical minimalist biomimetic computational panoramic imaging lens has the advantages of minimalist optical architecture, ultra-compact appearance, ultra-lightweight, and a large imaging area. 
As shown in Fig.~\ref{fig:Comparison}a, the traditional panoramic annular lens has the common optical architecture of larger panoramic lens blocks and relay lens groups. 
When the maximum FoV is large, due to the small thickness of the curved lens edge of the panoramic lens block, the thickness of the lens edge will be reduced again when increasing the effective aperture during processing, causing the edge of the curved lens to break or it is difficult to ensure that its edge FoV is not obstructed by external structures during manufacturing. 
Therefore, the actual imaging circle captured by cold processing technology is often smaller than the design value. The panoramic annular lens shown in the figure has 10 lenses and 1 infrared filter, with a total length of 79 mm and a maximum lens diameter of 44 mm. The designed maximum crescent lens has an edge thickness of 3.7 mm, and the ratio of edge thickness to diameter is less than 1/10. 
We installed the panoramic annular lens on the Sony A6600 and calculated the actual imaging circle to be 12.4 mm based on the actual captured images output by the Sony A6600~\cite{jiang2024minimalist}. 
According to the panoramic annular lens design, the actual manufacturing half-FoV can be inferred to be 92{\textdegree}. 
The imaging circle diameter of the spherical minimalist biomimetic computational panoramic imaging lens design is 10 mm, with an edge thickness of 2.8 mm, and the ratio of edge thickness to diameter is greater than 1/10. 
Due to the need to reserve a more effective aperture at the edges during manufacturing, the actual imaging circle diameter obtained from the calculated images is 10.4 mm. Therefore, in the case where the imaging target surfaces of the two panoramic lenses are relatively close, our proposed spherical minimalist biomimetic computational panoramic imaging lens breaks the optical architecture of traditional panoramic annular lenses that require at least 7-11 lenses. 

Based on the visual imaging principle of bionic four-eyed fish, we innovatively propose a minimalist panoramic lens that realizes super-large FoV panoramic imaging by using only a single lens and can flexibly and efficiently increase the edge thickness and effective diameter.
The prototype of the manufactured panoramic lens meets the design specification of 95{\textdegree} imaging HFoV. Combined with computational imaging, the system proposed in this paper can achieve better imaging perception capability. In terms of the size of the panoramic optical system, our proposed spherical minimalist biomimetic computational panoramic imaging lens reduces the diameter by 38\% and the total length by 82\%, achieving an ultra-compact panoramic imaging optical architecture. As shown in Fig.~\ref{fig:Comparison}b, the height of the traditional panoramic annular lens is 66 mm, and the weight is 157.5 g. The lens with a lens mount has a height of 17 mm and a weight of 29.5 g, while the lens system with only a lens has a height of 11 mm and a weight of 19.8 g. Therefore, our proposed biomimetic computational panoramic imaging system reduces the weight of the panoramic lens by more than 81\% in the weight of optical elements and lens barrel, while eliminating the problem of traditional panoramic annular lenses being difficult to process due to the large diameter and small edge thickness of the panoramic lens block. It has the ability of ultra-wide angle perception under large target imaging with a simple system, small size, light weight, and low cost. 
In the comparison of lens diameters shown in Fig.~\ref{fig:Comparison}c, it can be seen that our spherical minimalist biomimetic computational panoramic imaging lens still has the advantage of miniaturization, with only the lens size equivalent to a coin. Although the lens mount has a certain impact on the diameter and weight of the prototype of the minimalist biomimetic panoramic imaging lens, it is not limited to the design scheme of the metal structural component thread-fastening glass lens proposed in this paper. It can also use integrated plastic barrels and lens design schemes that use plastic as optical material to reduce the volume, weight, and cost of panoramic lenses (details in the appendix).

\subsection{Computational imaging method for enhanced panoramic imaging quality}
\label{sec:ci}
\begin{figure*}[!t]
  \centering
  \includegraphics[width=0.85\linewidth]{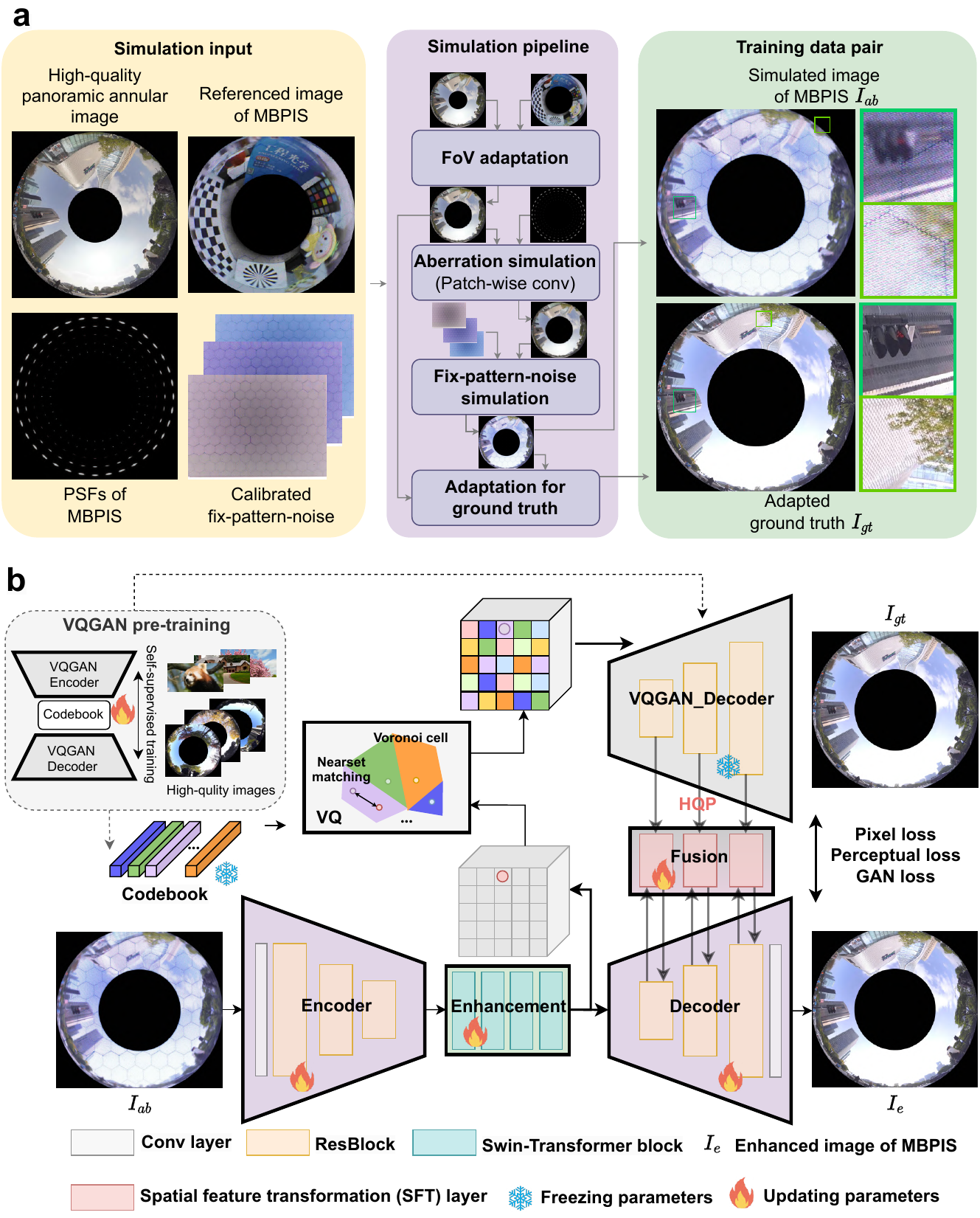}
  \caption{\textbf{Overview of the proposed deep-learning-based computational imaging method for enhancing imaging quality of MBPIS.} \textbf{a} Dataset construction method. We simulate the compound image degradation induced by MBPIS, \textit{i.e.}, the optical degradation caused by aberrations, and the fix-pattern noise caused by the micro-channel glass array sensor, on the high-quality PAL images to generate training data pairs. \textbf{b} Prior-embedded Image Enhancement Network (PIENet). We introduce the High-Quality Priors (HQP) learned by a VQGAN to compensate for the loss of image information induced by the severe degradation, modeling image enhancement as a generative paradigm via Vector Quantization (VQ).}
  \label{fig:cac}
\end{figure*}

As an optical design trade-off between minimalist optical system architecture and imaging quality, the imaging quality of MBPIS exhibits certain deficiencies, which are reflected in a compound form of image degradation.
To be specific, two types of image degradation dominate: one is the uncorrected optical aberration caused by the absence of relay lens groups, and the other is the fix-pattern noise introduced by the optical transmission in micro-channel glass array.
Such degradation brings significant information loss to the MBPIS images, limiting its convincing downstream applications.
Aiming at recovering the lost image details from the compound degradation, we propose a deep-learning-based computational imaging method to enhance the MBPIS images as shown in Fig.~\ref{fig:cac}, where an image enhancement model is pre-trained to generate high-quality panoramic images.

The key to training the model is to acquire the aligned, degraded-clear data pairs of MBPIS.
To this end, we first put forward a dataset construction method based on simulating the compound degradation of MBPIS on an open-resource, high-quality panoramic annular image dataset PALHQ~\cite{jiang2024minimalist} (details in the appendix). 
As shown in Fig.~\ref{fig:cac}a, the optical degradation is simulated via patch-wise convolution based on the PSFs of MBPIS calculated by ray tracing, while the noise pattern of the micro-channel glass array sensor is experimentally calibrated for fix-pattern noise simulation on the aberration image.
For simulating more realistic MBPIS images, our simulation pipeline also takes into account the differences in the FoVs and blind areas between referenced MBPIS images and high-quality images from PALHQ. 
In addition, to compensate for the changes in image brightness and color caused by fix-pattern noise simulation, we accordingly conduct adaptation for the ground-truth images, alleviating the impacts of these factors on model training. 
The detailed operations for each step of the simulation pipeline can be found in the appendix.

With the data pairs of simulated images of MBPIS $I_{ab}$ and adapted ground truth $I_{gt}$, we can train a deep network for image enhancement via supervised learning. 
However, as seen on the right side of Fig.~\ref{fig:cac}a, the degradation of MBPIS images leads to a severe loss of image textures, which poses a challenge for image enhancement networks trained solely with conventional pixel-wise objective~\cite{jiang2024minimalist}.
To address this limitation, as shown in Fig.~\ref{fig:cac}b, we build up a generative image enhancement paradigm by introducing the High-Quality Priors (HQP) learned by a VQGAN~\cite{van2017neural}, which stores the key features for reconstructing the high-quality images into a codebook and an image decoder. 
Differing from other work of VQGAN-based image enhancement, we specifically incorporate images from PALHQ~\cite{jiang2024minimalist} to further fine-tune VQGAN for MBPIS. 
In this way, the codebook in VQGAN contains feature vectors of high-quality panoramic annular images, based on which the decoder is also capable of generating corresponding high-quality features.
The Prior-embedded Image Enhancement Network (PIENet) is then designed to match the high-quality features in HQP with the input MBPIS image, enhancing the MBPIS image guided by the VQGAN decoder fed with the matched features.
Concretely, the deep features of the MBPIS image after the encoder are pre-enhanced by the Swin-Transformer-based~\cite{liang2021swinir} enhancement module to ensure convincing feature matching with the pre-learned codebook, which is then fed to the frozen VQGAN decoder to generate high-quality image features to guide the main enhancement branch through the Spatial Feature Transformation (SFT) layer~\cite{Wang_2018_CVPR} based fusion module.  
The training objective for PIENet is the combination of pixel loss, perceptual loss, and GAN loss, targeting at production of enhanced images with both high fidelity and visually pleasant image details (details in the appendix).

\begin{figure*}[!t]
  \centering
  \includegraphics[width=1.0\linewidth]{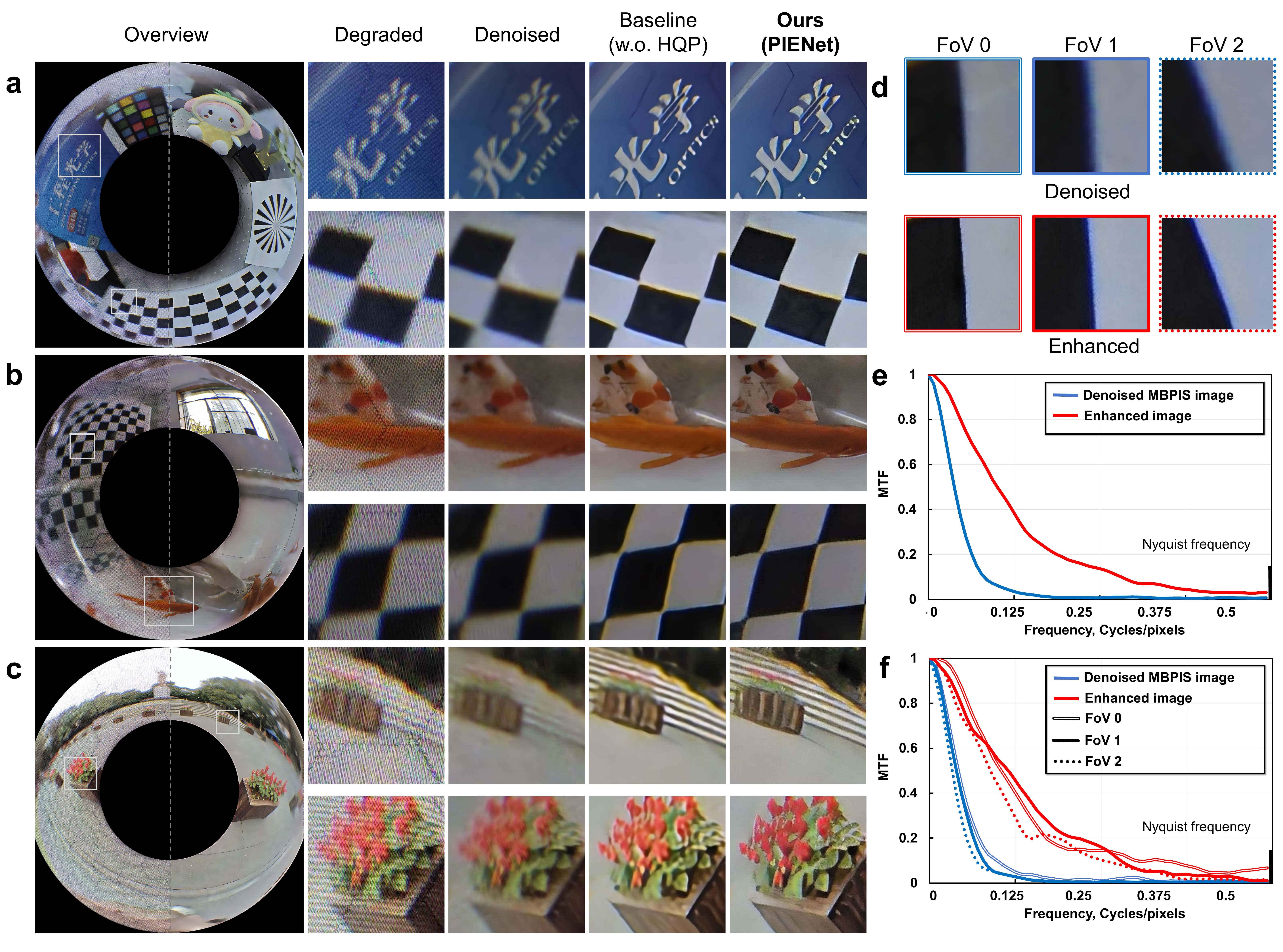}
  \caption{\textbf{Experimental results of computational imaging.} We compare the proposed PIENet with the state-of-the-art baseline U-Net equipped with both CNN and Transformers. The denoised results are also presented as a reference for demonstrating the optical aberrations of the MBPIS. Three distinct testing scenes are provided. \textbf{a} An indoor scene of close objects. \textbf{b} An amphibian visual imaging scene. \textbf{c} An outdoor open scene to showcase the generalization of our minimalist biomimetic computational panoramic imaging system. We also provide the results of MTF enhancement. \textbf{d} The ``knife edge'' image patches of different FoVs used for spatial frequency response (SFR) testing. \textbf{e} Average MTF enhancement results over all selected FoVs. \textbf{f} Detailed MTF enhancement results of each selected FoV. For MTF calculation of the original MBPIS images, SFR testing is conducted on the denoised images to avoid the influence of fixed-pattern noise on edge sampling.}
  \label{fig:visual}
\end{figure*}

The experimental results of the proposed computational imaging method are shown in Fig.~\ref{fig:visual}, including the visual results under three different scenes (Fig.~\ref{fig:visual}a${\sim}$c) and the MTF enhancement results (Fig.~\ref{fig:visual}d${\sim}$f) measured by Spatial Frequency Response (SFR) testing.
The deep learning network trained on our simulated data pairs $\{I_{ab}, I_{gt}\}$ can effectively address the issues of fix-pattern noise and optical degradation in the original MBPIS degraded images, contributing to the production of high-quality enhanced images with almost no residual fix-pattern noise and optical aberrations induced image blur.
However, due to the severe information loss caused by this compound degradation, the baseline method can hardly recover the lost details and textures. 
In this case, the proposed PIENet with the introduction of HQP can generate more realistic and richer image details, effectively addressing the image degradation in MBPIS.
Moreover, the excellent performance of the computational imaging method in indoor, amphibious, and outdoor scenes demonstrates the broad application prospects of the designed system, meeting the demands for minimalist and high-quality panoramic imaging.
In addition, the results of MTF enhancement further demonstrate the effectiveness of our method in correcting optical degradation. 
The proposed computational imaging method can nearly double the area under the MTF curve in both cases of overall and sub-FoVs SFR testing. 
The ``knife edge'' images in Fig.~\ref{fig:visual}d also indicate that the enhanced results are sharper with higher contrast.
More experiments and experimental settings can be found in the appendix. 

\section{Method and Conclusion}
\begin{figure*}[!t]
  \centering
  \includegraphics[width=1.0\linewidth]{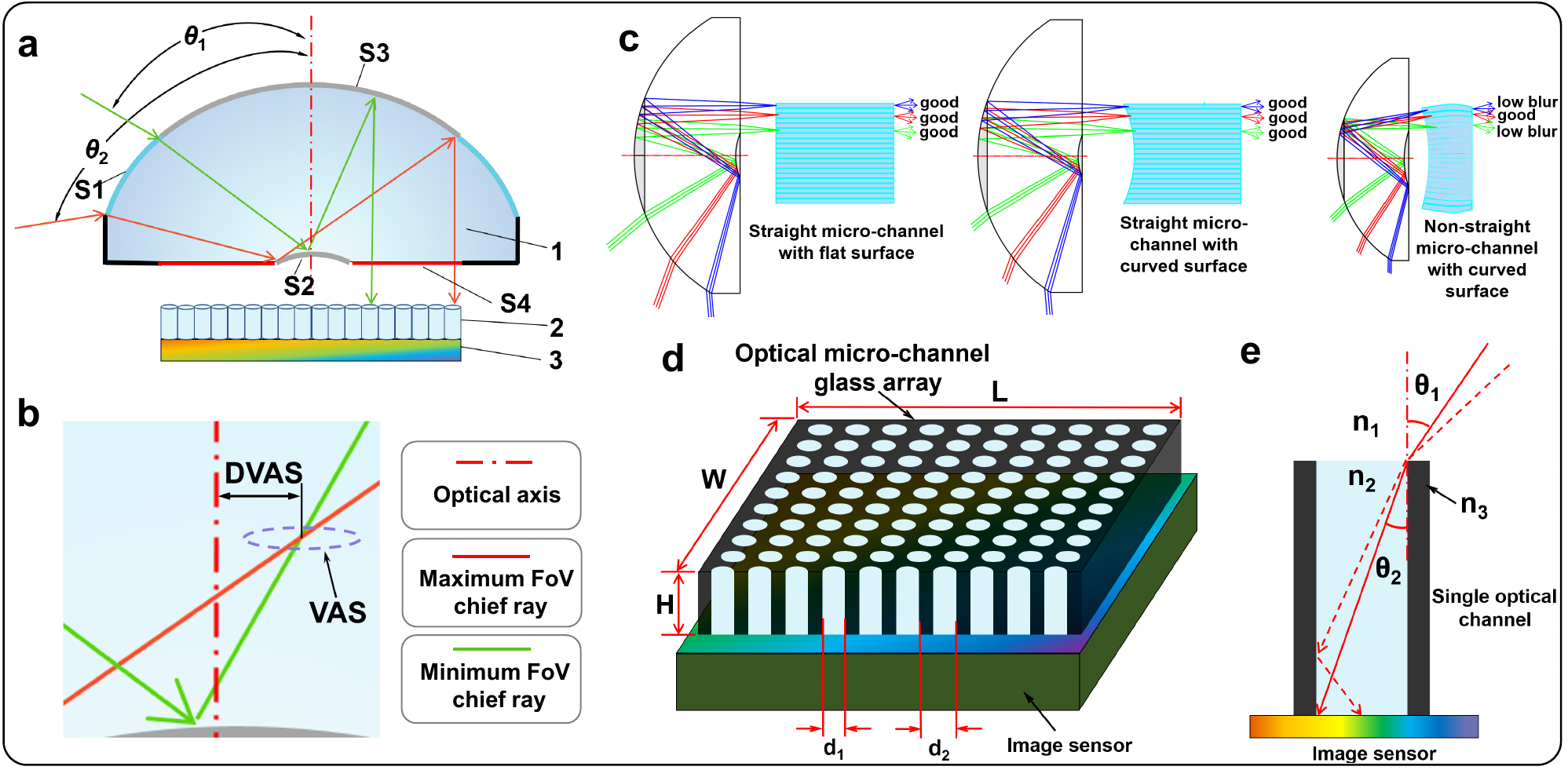}
  \caption{\textbf{Imaging architecture and optical micro-channel scheme.} \textbf{a} Imaging principle. \textbf{b} Off-axis virtual aperture design. \textbf{c} Design degree of freedom changes brought by three different micro-channel components. \textbf{d} Optical micro-channel glass array image sensor. \textbf{e} Single channel ray constrained imaging principle.}
  \label{fig:microchannel}
\end{figure*}

The imaging principle of this biomimetic-inspired minimalist panoramic system is shown in Fig.~\ref{fig:microchannel}a, where 1 is the panoramic lens, 2 is the optical micro-channel glass array, and 3 is the image sensor. The micro-channel array is coupled to the image sensor, and the panoramic lens is threaded and adjusted for back focus through the lens mount and sensor mount. The red dashed line represents the optical axis, the minimum imaging half FoV is $\theta_1$, the maximum imaging half FoV is $\theta_2$, the green solid line represents the incident light at the minimum FoV, and the red solid line represents the incident light at the maximum FoV. The incident light sequentially passes through S1 coated with anti-reflective film, S2 coated with reflective film, S3 coated with reflective film, and S4 coated with infrared cutoff film before entering the micro-channel array and image sensor. The surfaces of the remaining lenses are painted black to absorb stray light. Among them, S1 and S3 have the same radius. The line connecting the surface vertex of S2 to the center of curvature points perpendicular to the sensor, while the surface of S4 is flat. When designing this minimalist panoramic system, it is necessary to achieve as much as possible the image side telecentric optical path, that is, the incident angle of the principal ray on the image plane is 0{\textdegree} so that the same numerical aperture micro-channel array can be used to constrain the imaging beams of different FoVs. When designing this special reflective system, due to the lack of a true aperture stop that traditional panoramic systems have, this system has a minimalist imaging architecture. 

This design introduces a new off-axis virtual aperture for design, which has higher design freedom, as shown in Fig.~\ref{fig:microchannel}b. Since each imaging beam only requires its image centroid, the virtual aperture can become a novel design variable for aberration optimization of the optical path. This design method of using a virtual aperture inside the lens does not require additional complexity and volume of the system and is more advantageous compared to traditional panoramic imaging systems. In Fig.~\ref{fig:microchannel}c, we demonstrate three different micro-channel schemes, which can use fiber bundles, fiber optic panels, or optical micro-channel glass arrays. The left side scheme is a common planar micro-channel scheme with a planar end face, which can transfer the planar image plane to the planar image plane. The middle solution is to increase the optimization design variables so that the micro-channel end face becomes a curved surface with a radius to match the Petzval field curvature on the image plane brought by large field panoramic imaging. This approach can achieve better imaging quality, but it increases the polishing process for the micro-channel end face. The solution on the right is to use non-straight micro-channels for design, which eliminates the need to design the imaging beam of the optical system into an imaging telecentric shape. It can reduce the difficulty of chief ray angle (CRA) for different FoVs, greatly reducing system volume and size. However, due to the inconsistent magnitude of aberrations in each FoV, it is difficult to unify the CRA of each FoV. Instead, when optimizing the curved micro-channel end face to a certain radius, the aberrations of all FoVs are balanced. The design of this scheme poses difficulties in manufacturing and positioning due to the one-to-one coupling between the image end face of the micro-channel and the end face of the sensor.

The optical micro-channel glass array has the advantages of solid structure, high strength, easy cleaning, great light collimation performance, and good processability. This makes the optical micro-channel glass array more durable and maintainable than fiber bundles. 
Compared with traditional fiber bundles and fiber optic panels, the thickness of optical micro-channel glass arrays can reach a level of sub-millimeter, with low cost and extremely thin size, resulting in less transmission loss and easy mass production and application. This low-cost micro-channel solution brings enormous potential for the widespread application of biomimetic minimalist computing panoramic systems. To reduce the requirements for manufacturing optical micro-channels, we adopted a straight optical micro-channel scheme with a planar end face as shown in Fig.~\ref{fig:microchannel}c, and coupled an optical micro-channel glass array made of solid glass substrate material to the image sensor as shown in Fig.~\ref{fig:microchannel}d, forming a new type of optical micro-channel sensor for imaging. The optical micro-channel glass array panel is composed of a regularly arranged solid and transparent array of optical micro-channels. The substrate material is an absorbing glass with a refractive index similar to that of the micro-channels, with a length of L, a width of W, and a height of H. The diameter of a single channel is $\mathrm{d}_1$, and the distance between a single channel and another channel is $\mathrm{d}_2$. 
As shown in Fig.~\ref{fig:microchannel}e, it is the constraint of a single channel on the numerical aperture of the image beam. The incident light with an aperture angle smaller than that of a single channel is emitted from the transparent glass channel, while the light with an aperture angle larger than that is absorbed by the black glass, thereby achieving a numerical aperture constraint on the incident light. 
Due to the material of the substrate glass being a light-absorbing material, it can absorb stray light from all directions, making the image sharper and clearer. The refractive index of the substrate glass of the micro-channel glass array is lower than that of the glass material of the optical micro-channel. 
Light (non-imaging beam) with an aperture angle greater than $\theta_1$ will undergo multiple total internal reflections in the micro-channel glass medium and eventually be absorbed by the substrate glass due to not meeting the total internal reflection conditions, achieving the goal of eliminating stray light at the hardware level. When the thickness H of the micro-channel glass array is relatively thin, due to the fewer total internal reflections, the aperture angle of the incident light that can ultimately propagate to the image sensor is usually greater than the aperture angle $\theta_1$.

\begin{figure*}[!t]
  \centering
  \includegraphics[width=1.0\linewidth]{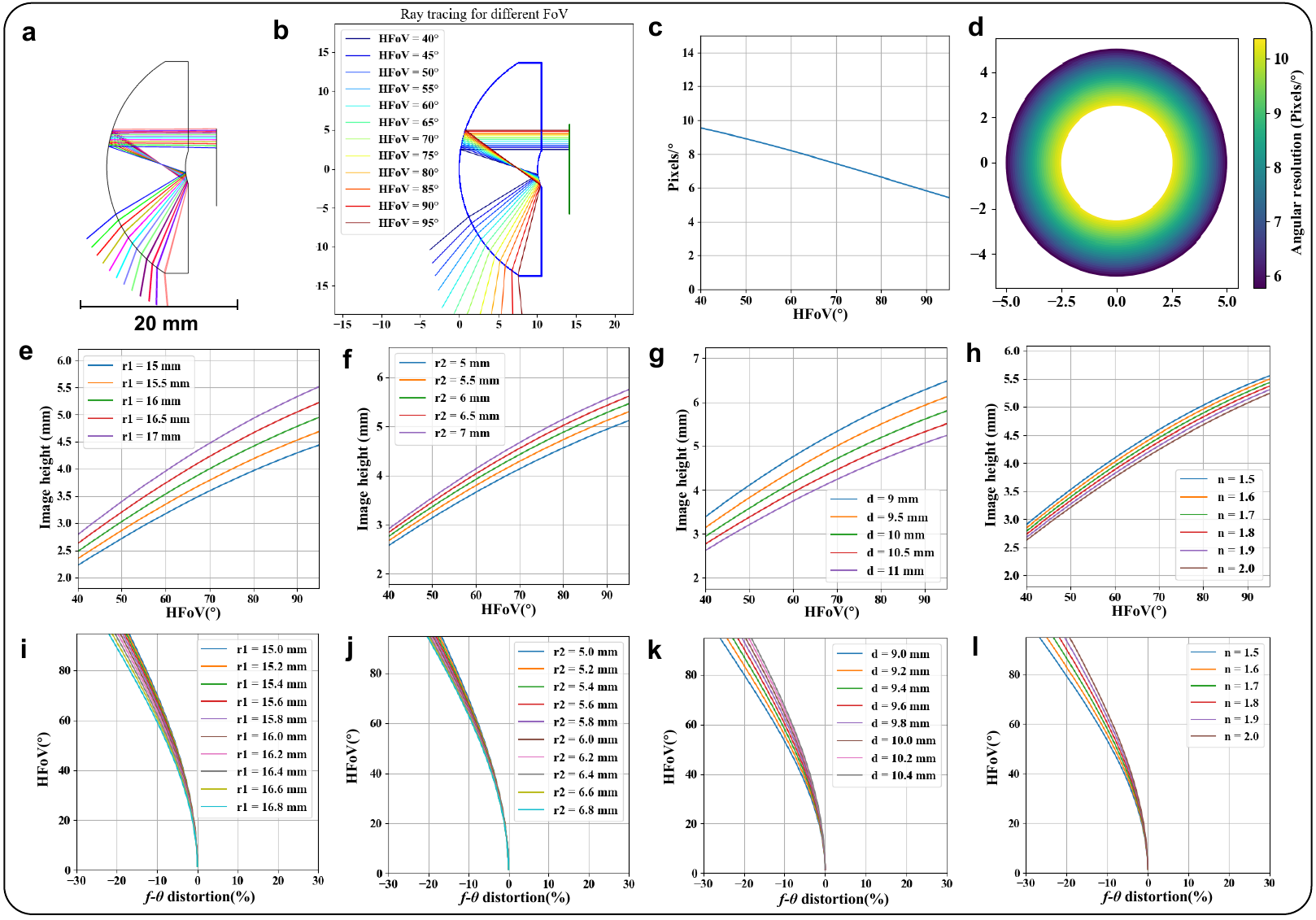}
  \caption{\textbf{Inverse ray tracing model and physical parameter relationships for a minimalist bionic computational panoramic imaging system.} \textbf{a} Minimalist bionic computational panoramic imaging system in a conventional optically designed light path. \textbf{b} Inverse ray tracing model. \textbf{c} Numerical curves of angular resolution distribution. \textbf{d} Angular resolution image plane distribution. \textbf{e} The numerical relationship between half-FoV and image height with respect to the radius of curvature $\mathrm{r}_1$. \textbf{f} The numerical relationship between half-FoV and image height with respect to the radius of curvature $\mathrm{r}_2$. \textbf{g} The numerical relationship between the half-FoV and the image height with respect to the thickness d. \textbf{h} Numerical relationship between the half-FoV and the image height with respect to the refractive index n. \textbf{i} Numerical relationship between $f-\theta$ aberration and half-FoV with respect to radius of curvature $\mathrm{r}_1$. \textbf{j} Numerical relationship between $f-\theta$ aberration and half-FoV with respect to radius of curvature $\mathrm{r}_2$. \textbf{k} Numerical relationship between $f-\theta$ aberration and half-FoV with respect to thickness d. \textbf{l} Numerical relationship between $f-\theta$ aberration and half-FoV with respect to refractive index n.}
  \label{fig:raytrace}
\end{figure*}

\begin{figure*}[!t]
  \centering
  \includegraphics[width=1.0\linewidth]{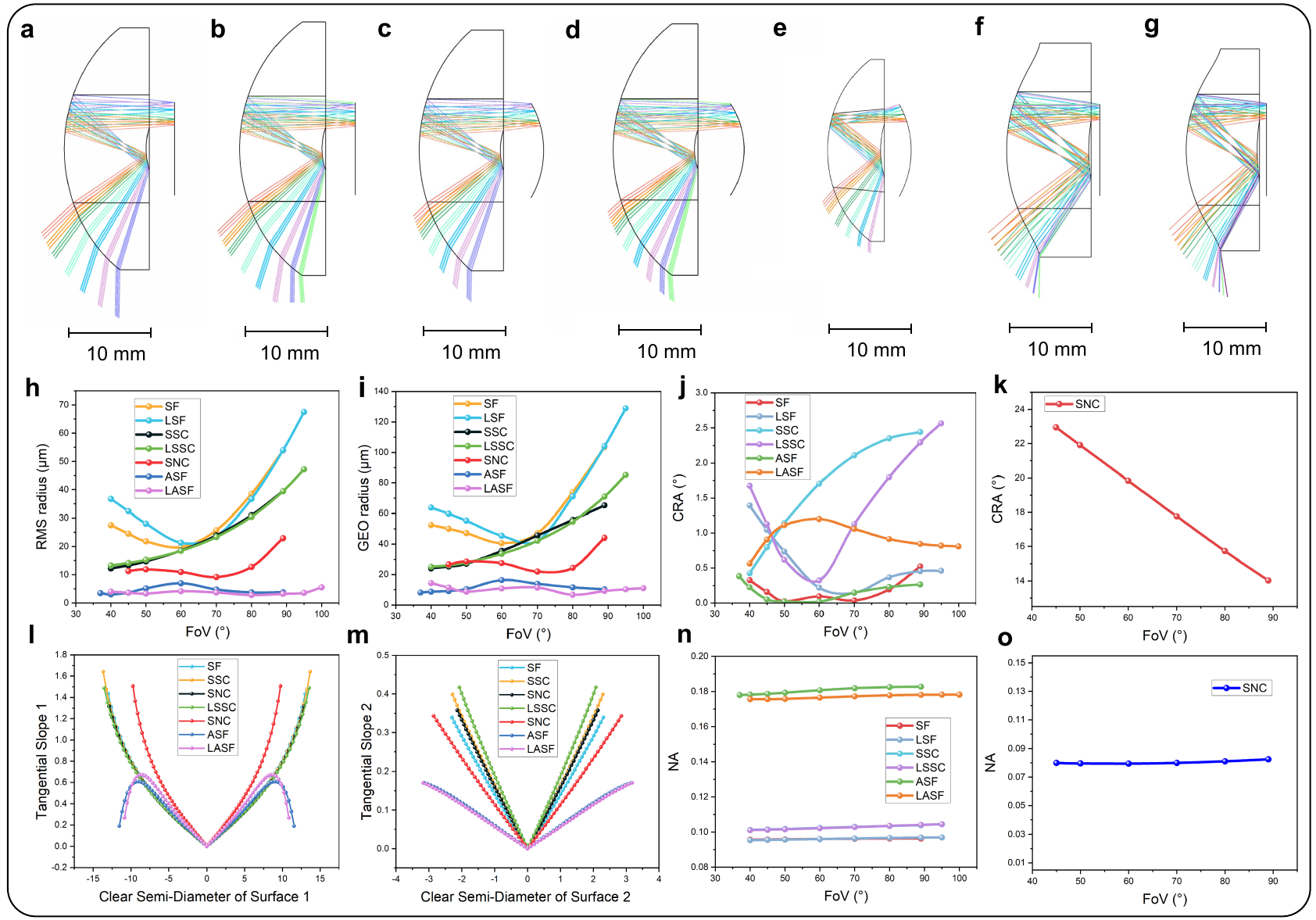}
  \caption{\textbf{Layout and imaging performance parameters comparison of 7 specifications of minimalist biomimetic panoramic computational imaging systems.} \textbf{a} The layout diagram of SF. \textbf{b} The layout diagram of LSF. \textbf{c} The layout diagram of SSC. \textbf{d} The layout diagram of LSSC. \textbf{e} The layout diagram of SNC. \textbf{f} The layout diagram of ASF. \textbf{g} The layout diagram of LASF. \textbf{h} RMS radius. \textbf{i} GEO radius. \textbf{j} CRA. \textbf{k} CRA of SNC. \textbf{l} The tangential slope of S1. \textbf{m} The tangential slope of S2. \textbf{n} NA. \textbf{o} NA of SNC.}
  \label{fig:Optical Architecture}
\end{figure*}

To further explore the imaging rules of this minimalist panoramic architecture, we established a ray tracing model for the imaging optical path of the system. To achieve a better coupling effect with the optical micro-channel array, the single lens reflection imaging optical path needs to be designed to be as far away from the center of the image as possible. Therefore, our ray tracing model is validated using the principal ray image center of each FoV. To verify the correctness of the modeling, we used traditional optical design software to validate the optical path of our minimalist panoramic architecture ray tracing model. As shown in Fig. \ref{fig:raytrace}a${\sim}$b, the results obtained from traditional commercial optical design software are completely consistent with those calculated using image side reverse ray tracing optical path, and the parameters of the incident and exit positions of the light rays are also completely consistent. To calculate the angular resolution, it is necessary to calculate the number of pixels occupied by each angle. For the convenience of manufacturing, we set the diameter $\mathrm{d}_1$ of a single micro-channel to 5 $\upmu$m and the micro-channel distance $\mathrm{d}_2$ to 6 $\upmu$m. The calculated angular resolution is shown in Fig. \ref{fig:raytrace}c. The angular resolution can be improved by reducing the diameter of channels and the spacing between channels. For example, the diameter $\mathrm{d}_1$ of a single micro-channel can be set to 3.5 $\upmu$m, and the distance $\mathrm{d}_2$ can be set to 4 $\upmu$m. It should be noted that the pixel size of the sensor is 1.6 $\upmu$m. To avoid Moire fringes caused by the coupling between the optical micro-channel glass array and the image sensor, the diameter of a single micro-channel should not be exactly the same as or too close to the sensor size. At the same time, in the process of coupling to form the optical micro-channel glass array sensor, it is necessary to view the coupling imaging effect displayed on the monitor in real time, so as to avoid the generation of Moire fringes. The angular resolution on the image plane obtained by reverse optical path tracing is shown in Fig. \ref{fig:raytrace}d, and the central blank area is the central blind spot. 
Fig.~\ref{fig:raytrace}e${\sim}$h shows the variation curves of the first refractive surface radius $\mathrm{r}_1$, first reflective surface $\mathrm{r}_2$, lens center thickness d, and lens refractive index n as variables with respect to the half-FoV and image height, respectively. It can be seen that $\mathrm{r}_1$ has the greatest impact on image height, while $\mathrm{r}_2$ has a relatively small impact on image height. As the thickness and refractive index increase, the image height gradually decreases. Therefore, when optimizing the design, the design weight can be mainly placed on the optimization of $\mathrm{r}_1$, while using smaller lens thickness and low refractive index optical materials can make it easier to achieve large target area imaging. Due to the large FoV of the panoramic system, it is necessary to focus on the physical parameters that have a significant impact on its $f-\theta$ distortion during design. Fig.~\ref{fig:raytrace}i${\sim}$l show the effects of four optimization variables, namely the radius $\mathrm{r}_1$ of the first refractive surface, the radius $\mathrm{r}_2$ of the first reflective surface, the thickness d, and the refractive index n, on the $f-\theta$ distortion of different FoVs. The larger the values of the first refractive surface radius $\mathrm{r}_1$ and the first reflective surface $\mathrm{r}_2$, the greater the $f-\theta$ distortion of the system. In addition, under the same amount of variation, the effect of $\mathrm{r}_1$ on $f-\theta$ distortion is greater than that of $\mathrm{r}_2$. The larger the thickness d at the center of the lens and the higher the refractive index, the smaller the distortion will be. Therefore, when designing this biomimetic minimalist computing panoramic lens, it is necessary to consider increasing the lens center thickness d to reduce distortion while balancing the control of image height and system volume. 
The use of high-refractive-index optical materials can significantly reduce system distortion, but it will increase the cost of the system. The optimization variables for standard spherical design are relatively small, so in order to control distortion, we use high refractive index glass for the design. When designing with aspherical surface types, we used low-refractive-index glass to reduce manufacturing costs. Therefore, we ultimately designed seven different minimalist biomimetic computing panoramic system schemes (as shown in Table~\ref{tab:2}). 
The optical architecture of the SNC system is shown in Figs.~\ref{fig:Optical Architecture}a${\sim}$g. 
The SNC system architecture can greatly reduce the system volume. To visually observe the influence of various system parameters on imaging performance indicators, we summarized the RMS radius, geometric radius, CRA, NA, and surface vector height of the point array diagram that vary with the FoV as a function of the net half aperture of the front surface S1 and rear surface S2 of the lens in Fig.~\ref{fig:Optical Architecture}h. Due to the special nature of the SNC system's non-straight optical channel imaging architecture compared to other architectures, the CRA and NA are presented separately. The parameter variation trend of the RMS radius and GEO radius of the point chart with respect to the FoV angle is very close, but not exactly the same. Aspherical surfaces can bring a large number of optimization variables to the system. The design schemes of ASF and LASF relative to spherical surfaces can greatly reduce the system volume, and high-performance imaging can be achieved with optical materials with a low refractive index of 1.45. 
The CRA changes of the SSC and LSSC schemes with curved ports in optical micro-channels are more drastic, but compared to the SNC scheme, the CRA still has the advantage of being far from the center of the image. The aspherical schemes, ASF and LASF, on the front surface S1 exhibit more significant changes in vector height compared to the rear surface S2. The NA of SNC is relatively smaller compared to other schemes, while the NA of other schemes does not change much in terms of FoV. All the proposed design schemes in this paper only explore the off-axis eccentricity of the virtual aperture as a variable. 
If it is desired to increase design freedom to optimize aberration design,  both the tilt and eccentricity of the virtual aperture can be set as variables. To investigate the influence of large FoV on size parameters and aberrations, seven different design schemes were explored for both normal imaging FoV and large FoV imaging scenarios. The above 7 optical design schemes all use glass materials. If glass materials are replaced with low-refractive-index plastic materials, the use of injection molding technology can bring promising low-cost solutions for mass production and application.

\begin{table*}[!ht]
\centering
\scriptsize
\begin{threeparttable}
\caption{Comparison of parameters for different panoramic imaging system architectures.\tnote{\textit{a}}}
\setlength{\tabcolsep}{0.5mm}{
\begin{tabular}{cccccccccccccc}
\toprule
\makecell[c]{Type
}&\makecell[c]{FL
(mm)
}& \makecell[c] {HFoV
($^{\circ}$)}&\makecell[c]{F/\#
}&\makecell[c]{SR ($\upmu$m)}&\makecell[c]{MTF (lp/mm)}&\makecell[c]{D ($\%$)}&\makecell[c]{RI}&\makecell[c]{RIP (mm)}&\makecell[c]{DMT (mm)}&\makecell[c]{TL (mm)}&\makecell[c]{IHMFV (mm)}&\makecell[c]{ET (mm)}\\
\midrule
\makecell[c]{SF}&\makecell[c]{3.9
}&\makecell[c]{40 $\sim $89}&\makecell[c]{5.3}&\makecell[c]{27$\sim $54}&\makecell[c]{Cutoff@7
}&\makecell[c]{-21.3}&\makecell[c]{0.98}&\makecell[c]{inf}&\makecell[c]{26}&\makecell[c]{13.6}&\makecell[c]{2.61}&\makecell[c]{3.6}\\
\midrule
\makecell[c]{LSF}&\makecell[c]{3.7
}&\makecell[c]{40 $\sim $95}&\makecell[c]{5.3}&\makecell[c]{36$\sim $67}&\makecell[c]{Cutoff@5
}&\makecell[c]{-23.2}&\makecell[c]{0.98}&\makecell[c]{inf}&\makecell[c]{27.3}&\makecell[c]{14.1}&\makecell[c]{2.49}&\makecell[c]{2.8}\\
\midrule
\makecell[c]{SSC}&\makecell[c]{3.9
}&\makecell[c]{40 $\sim $89}&\makecell[c]{5.0}&\makecell[c]{12$\sim $39}&\makecell[c]{Cutoff@12
}&\makecell[c]{-18.3}&\makecell[c]{0.95}&\makecell[c]{9}&\makecell[c]{25.7}&\makecell[c]{15.2}&\makecell[c]{2.61}&\makecell[c]{3.9}\\
 \midrule
\makecell[c]{LSSC}&\makecell[c]{3.7
}&\makecell[c]{40 $\sim $95}&\makecell[c]{5.0}&\makecell[c]{13$\sim $47}&\makecell[c]{Cutoff@9
}&\makecell[c]{-19.8}&\makecell[c]{0.92}&\makecell[c]{8}&\makecell[c]{27.0}&\makecell[c]{16.1}&\makecell[c]{2.46}&\makecell[c]{3.1}\\
\midrule
\makecell[c]{SNC}&\makecell[c]{4.2
}&\makecell[c]{45 $\sim $89}&\makecell[c]{5.7}&\makecell[c]{9$\sim $23}&\makecell[c]{Cutoff@17
}&\makecell[c]{-38.9}&\makecell[c]{0.85}&\makecell[c]{9}&\makecell[c]{19.4}&\makecell[c]{10.1}&\makecell[c]{3.07}&\makecell[c]{1.6}\\
\midrule
\makecell[c]{ASF}&\makecell[c]{5.4
}&\makecell[c]{37 $\sim $89}&\makecell[c]{2.8}&\makecell[c]{3$\sim $7}&\makecell[c]{0.13@100
}&\makecell[c]{-41.2}&\makecell[c]{0.96}&\makecell[c]{inf}&\makecell[c]{23.1}&\makecell[c]{11.5}&\makecell[c]{3.30}&\makecell[c]{6.3}\\
\midrule
\makecell[c]{LASF}&\makecell[c]{5.1
}&\makecell[c]{40 $\sim $100}&\makecell[c]{2.8}&\makecell[c]{3$\sim $5}&\makecell[c]{0.20@100
}&\makecell[c]{-44.3}&\makecell[c]{0.97}&\makecell[c]{inf}&\makecell[c]{21.7}&\makecell[c]{9.8}&\makecell[c]{3.36}&\makecell[c]{4.8}\\

\bottomrule
\label{tab:2}
\end{tabular}}

\begin{tablenotes}
        \footnotesize
        \item[\textit{a}] SF: Spheric lens and Flat micro-channel. LSF: Large FoV Spheric lens and Flat micro-channel. SSC: Spheric lens and Straight micro-channel with Curved end face. LSSC: Large FoV Spheric lens and Straight micro-channel with Curved end face. SNC: Spheric lens and Non-straight micro-channel with Curved end face. ASF: Aspheric lens and Flat micro-channel. LASF: Large FoV Aspheric lens and Flat micro-channel. FL: Focal Length. HFoV: Half Field of View. F/\#: F-number. SR: Spot root mean square (RMS) Radius. MTF: Modulation Transfer Function. D:  $f$-$\theta$ Distortion. RI: Relative Illumination. RIP: Radius of Image Plane. DMT: Diameter. TL: Total Length. IHMFV: Image Height of Minimum FoV. ET: Edge Thickness.
    
      \end{tablenotes}
      \small
    \end{threeparttable}
\end{table*}

Inspired by the principle of four-eyed fish visual imaging, we propose a biomimetic minimalist panoramic computing imaging system and creatively propose a novel optical micro-channel glass array sensor that can achieve hardware-level absorption of stray light while constraining the image space beam. This panoramic system has changed the complex architecture of traditional 7$\sim$11 lens panoramic systems, achieving ultra-wide FoV panoramic imaging with only a single ultra-simple lens. Compared with traditional panoramic annular lenses, our proposed biomimetic minimalist panoramic computing imaging system reduces the total length of the panoramic lens by 82\% and the weight by more than 81\%. 
At the same time, it has the advantages of being ultra-compact, lightweight, simple in structure, and low-cost. The biomimetic minimalist panoramic computing imaging system proposed in this paper breaks through the limitations of traditional panoramic lens optical architecture and solves the problems of complex structure, tedious processing technology, large volume, and heavy weight that traditional panoramic imaging systems cannot be applied to scenarios with strict volume and weight restrictions such as miniaturized drones and robots. By combining computational imaging technology, we have further enhanced its imaging perception capability. There are still some areas for improvement in the current system, such as using algorithms to eliminate image distortion. 
In the future, we aim to further improve this panoramic lens by constructing it into a plastic-shaped lens, designing and producing it using injection molding technology, achieving hardware-level low-distortion high-performance design, further reducing its volume and weight, and improving the imaging quality of its optical system. The biomimetic minimalist panoramic computing imaging system proposed in this paper brings great application prospects for high-quality and miniaturized panoramic perception optical systems and is expected to be widely used in many fields such as intelligent perception, security monitoring, pipeline detection, video conferencing, computer vision, \textit{etc.}

\subsection{Novel Optical Micro-channel Glass Array Sensors and System Components}
\label{sec:refs}

\begin{figure*}[!t]
  \centering
  \includegraphics[width=1.0\linewidth]{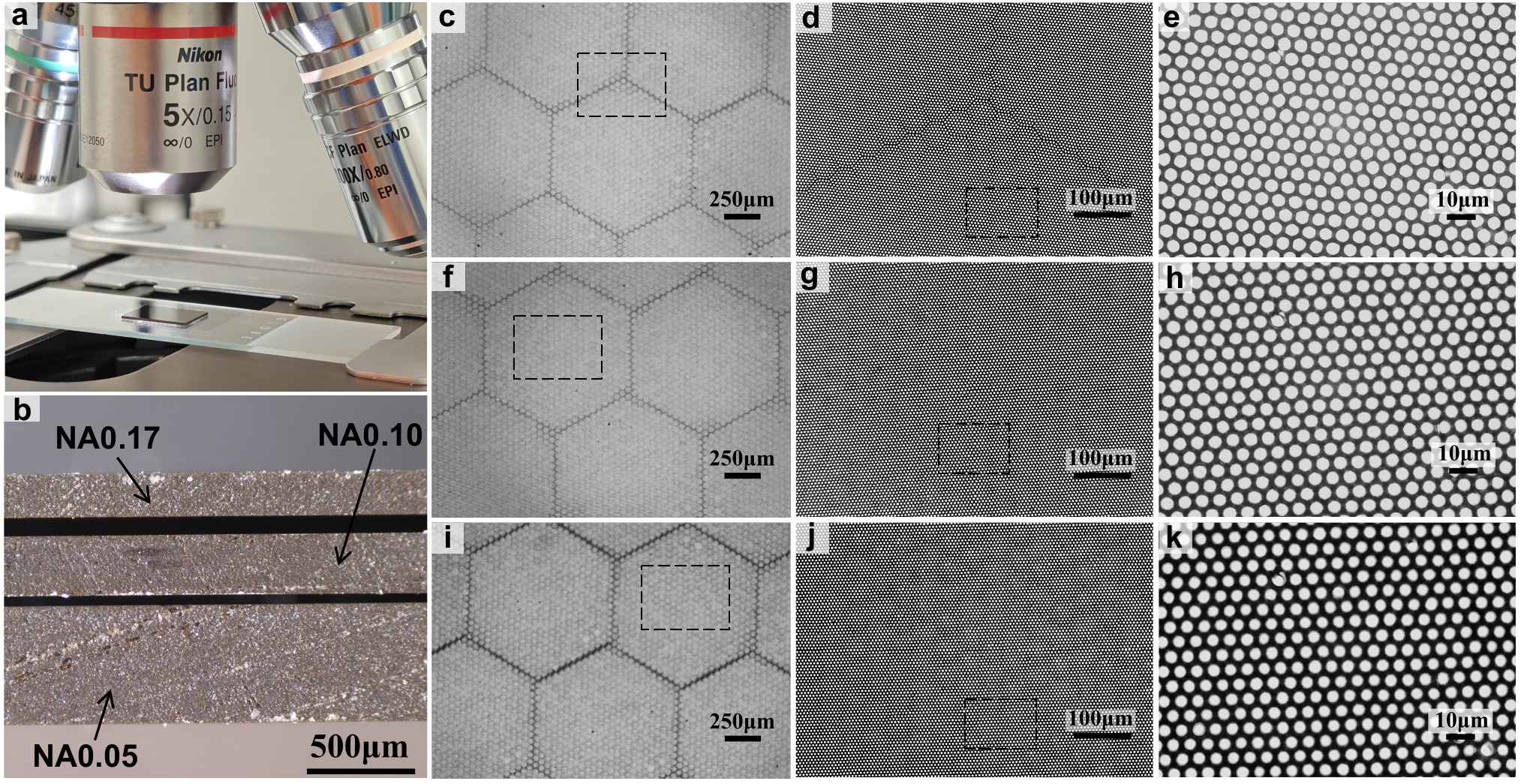}
  \caption{\textbf{Microscopic imaging results of surface microstructure and thickness of optical micro-channel glass arrays with three different NA specifications.} \textbf{a} Use microscopes with magnifications of 5$\times$, 20$\times$, and 100$\times$ to capture microstructure images of optical micro-channel glass arrays placed on glass slides at different scales. \textbf{b} Imaging results of three different thicknesses of optical micro-channel glass arrays under a 5$\times$ microscope. \textbf{c} NA0.17 under a 5$\times$ microscope. \textbf{d} NA0.17 under a 20$\times$ microscope. \textbf{e} NA0.17 under a 100$\times$ microscope. \textbf{f}  NA0.10 under a 5$\times$ microscope. \textbf{g} NA0.10 under a 20$\times$ microscope. \textbf{h} NA0.10 under a 100$\times$ microscope. \textbf{i} NA0.05 under a 5$\times$ microscope. \textbf{j} NA0.05 under a 20$\times$ microscope. \textbf{k} NA0.05 under a 100$\times$ microscope.}
  \label{fig:microstructure}
\end{figure*}

To validate our minimalist bionic computational panoramic lens scheme, we first fabricate optical micro-channel arrays with NAs of 0.17, 0.10, and 0.05, respectively. As shown in Fig.~\ref{fig:microstructure}, we placed the fabricated optical micro-channel glass arrays on slides and used an optical microscope with 5$\times$, 20$\times$, and 100$\times$ for microstructure inspection and imaging of the optical micro-channel glass arrays, respectively. 
Fig.~\ref{fig:microstructure}b shows the thickness difference of this kind of micro-channel glass arrays, with the theoretical fabricated thicknesses of 0.18 mm, 0.25 mm, and 0.5 mm for NA of 0.17, 0.10, and 0.05. 
Due to the use of the polishing technique to correct the fabricated thicknesses and the existence of certain tolerance in the processing, the actual processed thicknesses of the optical micro-channel glass arrays were 0.21 mm, 0.28 mm, and 0.52 mm, respectively. 
The increase in the thickness of the optical micro-channel glass arrays causes the actual transmission distance of the light traveled in the optical micro-channel plate to increase slightly compared to the theoretical fabrication thickness of the light traveled, which has an effect on the imaging of the system as a result of which the F/\# of the final system is slightly larger and the point spread function is slightly smaller, and thus the aberration of the optical system collected by the sensor will be smaller than that of the theoretical design. 

In Fig.~\ref{fig:microstructure}c, Fig.~\ref{fig:microstructure}f, and Fig.~\ref{fig:microstructure}i the optical boundaries formed by splicing ortho-hexagonal shapes with an NA of 0.17, 0.10, and 0.05 captured under the observation FoV of a 5$\times$ microscope are shown, respectively, where each ortho-hexagonal region is a secondary composite unit with a side length of about 530 $\upmu$m. The black dashed boxed region was imaged separately using a 20$\times$ microscope for imaging to obtain Fig.~\ref{fig:microstructure}d, Fig.~\ref{fig:microstructure}g, and Fig.~\ref{fig:microstructure}j. The primary composite units near the boundary are ortho-hexagonal with each ortho-hexagon having a side length of 25 $\upmu$m. This optical boundary is more pronounced around the secondary composite units, while the optical boundary of the primary composite units inside the secondary composite units is not pronounced. The areas of the dashed box were imaged using a 100$\times$ microscope, respectively, as shown in Fig.~\ref{fig:microstructure}e, Fig.~\ref{fig:microstructure}h, and Fig.~\ref{fig:microstructure}k. Under the same illumination and exposure conditions, the optical micro-channel glass array with an NA of 0.17 allows more light to pass through and thus can be coupled with the sensor to provide image-square light confinement to the fast imaging system for a smaller F/\#.
The optical micro-channel glass array with an NA of 0.05 optical micro-channel glass arrays have a thicker transmitted light path of the three sizes, so most light is absorbed, transmitting less light and imaging the least amount of light energy per channel into the microscope sensor. The diameter of the individual micro-channels is 5 $\upmu$m, and the distance between the micro-channels, d2, is 6 $\upmu$m. If it is desired to improve the angular resolution of the final panoramic imaging system, the diameter of the individual micro-channels can be machined to smaller dimensions, and the distance between the micro-channels can be reduced. If different system F/\# values need to be realized, the light can be constrained by setting different thicknesses of optical micro-channel glass arrays.

\subsection{Results and Conclusion}
The current panoramic imaging systems require the use of real apertures for imaging, hence the emergence of relay lens groups. The emergence of relay lens groups brings optimization variables to the correction of large FoV aberrations in panoramic systems, but increases the volume and length of the system. In addition, the traditional design method for achieving high-resolution panoramic imaging of large imaging targets is to scale the traditional complex architecture of panoramic imaging systems. This method leads to a significant increase in system volume and weight during large sensor imaging area, making it difficult to equip for extreme application scenarios with strict limitations on volume and weight. To solve the problem of difficulty in achieving high-resolution imaging of large target surfaces in current panoramic system optical architectures with small sizes, we innovatively propose a new panoramic architecture with a ``Virtual aperture stop'' based on biomimetic principles. We use microchannel array sensors as imaging photosensitive elements to construct a single-lens minimalist biomimetic computational panoramic imaging system. The system can achieve panoramic imaging using only one lens optical architecture, reducing the size and weight of the panoramic imaging system. Combined with computational imaging to enhance image quality and matching with a 1-inch image sensor, it achieves high-quality panoramic imaging with 3K resolution on a large target surface under a minimalist architecture.

This paper constructs a single lens minimalist biomimetic computational panoramic imaging system, proposes a single lens panoramic system with a ``Virtual aperture stop", and uses microchannel array sensors to achieve aperture constraints on each FoV. Based on the physical structural parameters of the single-lens panoramic imaging system, a mathematical model for reverse ray tracing was proposed, and its physical structure and imaging parameters were analyzed in detail. Based on the analysis results, different single lens panoramic imaging systems were designed, including a large aperture F/\# 2.8 aspherical panoramic system. We have manufactured and validated a spherical single lens panoramic system with an F/\# of 5.3, and combined it with computational imaging to enhance the quality of panoramic images. We have achieved high-quality panoramic imaging with 3K image resolution on a 1-inch large target surface. Compared with traditional panoramic ring imaging systems that are close to the target surface, this system reduces the optical total length by 82\% and the weight by 81\%. In the spatial frequency response test results, a quantitative evaluation of the enhancement of image quality in computational imaging has been completed, which can be applied in different scenarios. The effectiveness of the single lens minimalist biomimetic computational panoramic imaging system was further validated through the comparison of visual effects and downstream tasks.

\section{Discussion}
\subsection{Manufacturing of optical components}
The MBPIS prototype consists of three core optical components: the monolithic panoramic lens, the optical micro-channel glass array, and the image sensor. The single panoramic lens was fabricated from high-refractive-index glass via cold processing. The spherical LSF lens achieves a focal length of 3.7 mm, a weight of 19.8 g, and a total optical length of 14 mm. The optical micro-channel glass arrays with three numerical apertures (0.05, 0.10, 0.17) were fabricated on solid glass substrates and directly coupled to a 1-inch Sony IMX989 sensor, replacing both the protective glass and infrared filter of a conventional sensor. Comprehensive manufacturing details for the lens and micro-channel array are provided in the appendix.
Following component fabrication, the lens mount and chip base were machined from metal with a threaded interface for back-focus adjustment, accommodating both the image sensor and the optical micro-channel glass array sensor. Non-optical surfaces were coated with matte black paint for stray light suppression. For mass production, these mechanical components can be redesigned using plastic injection molding to further reduce weight and cost in the future. The complete assembled prototype is shown in Fig. 7. Detailed mechanical specifications and assembly procedures are described in the appendix.

\subsection{Aberration simulation}
In our imaging simulation pipeline, the degradation caused by aberrations is characterized by the energy dispersion of the Point Spread Function (PSF).
We use Zemax to compute the diffraction PSF using direct integration of the Huygens wavelets method, and then our simulation model applies patch-wise convolution between the FoV-adapted PALHQ image $I_S$ and PSF map to generate the aberrated image $I_A$. Specifically, $I_S$ is divided into patches and then each patch (denoted as $I^{patch}_{S}$) is convolved with its corresponding PSF:
\begin{equation}
\label{eq:method}
    I^{patch}_A \approx PSF \ast I^{patch}_{S},
\end{equation}
where $I^{patch}_A$ is the degraded image patch. After that, we stitch all degraded image patches to obtain the complete aberration image $I_A$ (details in the appendix).

\subsection{Computational imaging method}
Our computational imaging method aims to recover a clear image $I_{clear}$ from an MBPIS image $I_{ab}$ with composite degradation, which is a typical ill-posed problem. To address this issue, we propose using a deep learning-based image restoration model. Given the model architecture $f$ and its learnable parameters $\theta$, the problem is formulated as finding the optimal $\theta^{*}$ that minimize the loss function $\mathcal{L}$ (detailed in the appendix):
\begin{equation}
    \bm{\theta} ^{*} =\argmin_{\bm{\theta}} \sum_{i=1}^{N}{\mathcal{L}}(f_{\bm{\theta}}(I_{ab}),I_{gt}),
    \label{eq:loss function}
\end{equation}
where $I_{gt}$ is the ground-truth clear image of $I_{ab}$.
Based on the formulation of this problem, the core of our computational imaging method lies in preparing the paired $I_{ab}$ and $I_{gt}$ data for model training. 
To achieve this, we propose a data construction method based on optical simulation, utilizing publicly available high-quality panoramic images directly as $I_{gt}$. We then simulate the composite degradation corresponding to the MBPIS images on $I_{gt}$ to obtain synthetic $I_{ab}$ (For specific details, please refer to the appendix).

\subsection{Video object segmentation}

Once the clear images $I_{clear}$ are obtained by the ultra-compact bionic computational imaging system, they can be directly integrated into a diverse range of downstream tasks.
Here we concentrate on video object segmentation, a representative and widely-used task that aims to predict a set of binary masks $\mathcal{S}=\{s_t \}^{T_V}_{t=1}, s_t \in \mathbb{R}^{H\times W}$ from a given video $\mathcal{V}=\{I_t\}^{T_V}_{t=1}$ with $T_V$ frames and a referring expression.
We leverage SAM2, a robust zero-shot video object segmentation model, to accomplish video object segmentation.
Without the need for fine-tuning on domain-specific data, it can function flawlessly in previously unseen scenarios.
It accepts several positive or negative annotations within certain frames as the referring expression and subsequently generates a mask set, $\mathcal{S}$, for the segmented object.
To enable simultaneous segmentation of multiple objects, annotations with varying object labels, $\{0,1,\cdots,i \}$, are provided, yielding mask sets $\{\mathcal{S}_i\}$ for each corresponding object (details in the appendix).

\bibliographystyle{SELF-cas-model2-names}
\bibliography{sample}

\section*{Funding}
This work was supported by the National Natural Science Foundation of China (Grant No. 12174341 and No. 62473139), Henan Province Key R\&D Special Project (Grant No. 231111112700), the Hunan Provincial Research and Development Project (Grant No. 2025QK3019), the Open Research Project of the State Key Laboratory of Industrial Control Technology, China (Grant No. ICT2025B20), the State Key Laboratory of Autonomous Intelligent Unmanned Systems (the opening project number ZZKF2025-2-10).

\clearpage
\section{Imaging model of minimalist biomimetic panoramic imaging system}

\begin{figure*}[!h]
  \centering
  \includegraphics[width=0.78\linewidth]{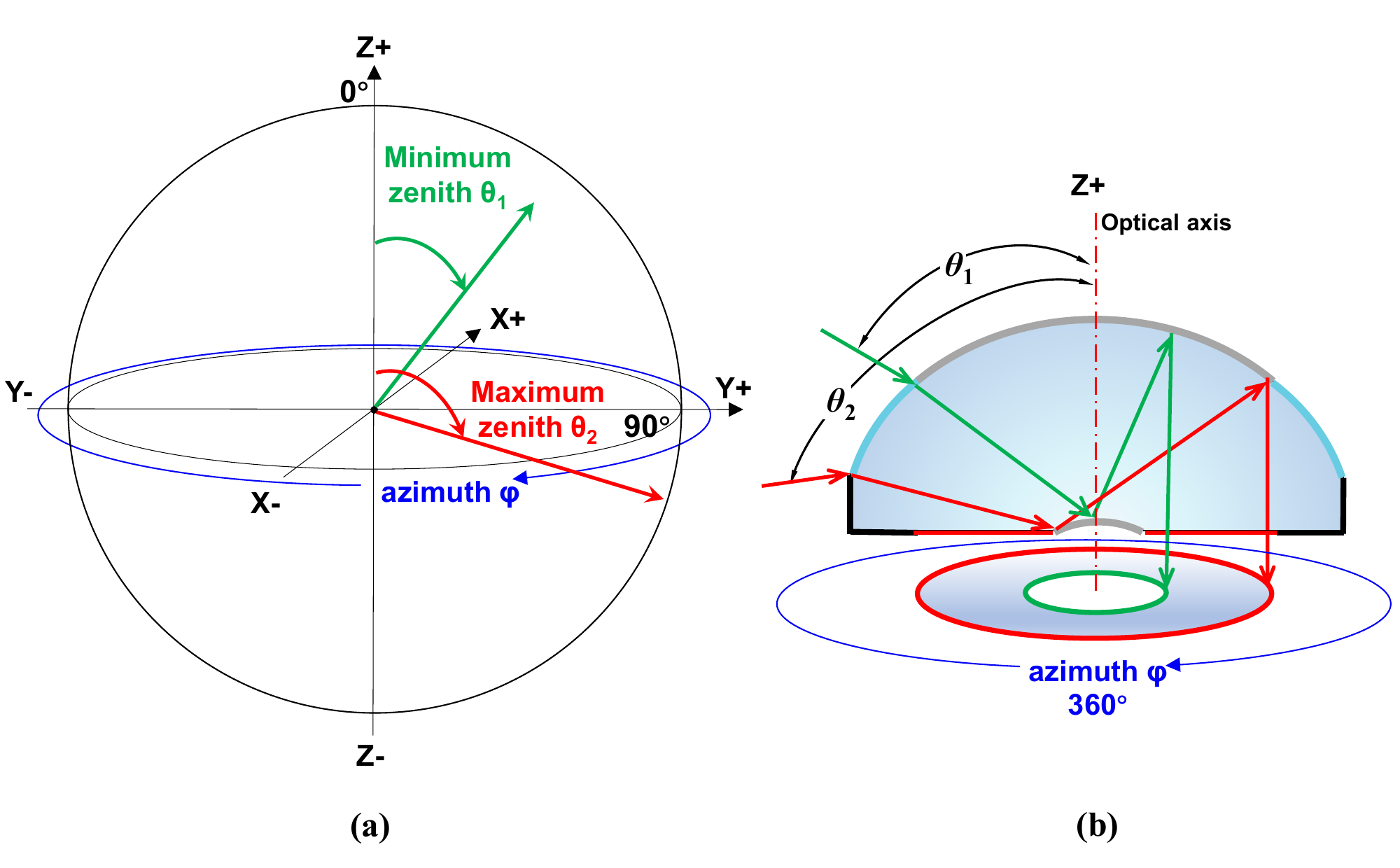}
  \caption{Spherical coordinate definition in MBPIS. (a) Three-dimensional view showing the azimuth angle $\phi$ (360° in the X–Y plane) and zenith angle $\theta$ (measured from the optical axis). (b) Cross-sectional view illustrating the imaging half FoV bounded by the minimum zenith $\theta_{1}$ and maximum zenith $\theta_{2}$, with the corresponding incident rays and the central blind zone.}
  \vskip-2ex
  \label{fig:azimuth and zenith}
\end{figure*}

The imaging model of the minimalist biomimetic panoramic imaging system (MBPIS) is established in a spherical coordinate system, as shown in Fig.~\ref{fig:azimuth and zenith}. The optical axis is aligned with the positive Z-direction, and the origin O coincides with the vertex of the front surface S1. The direction of each incident ray is parameterized by two angular coordinates: the azimuth angle $\phi$ and the zenith angle $\theta$. The azimuth $\phi$ spans the full 360° range in the X–Y plane, providing the system's panoramic perception capability. The zenith $\theta$ is measured from the optical axis (Z-axis). The annular imaging region on the sensor plane receives incident light within the zenith range from the minimum half field-of-view (FoV) $\theta_{1}$ to the maximum half FoV $\theta_{2}$. As described in the main text (Section 3), the ray at $\theta_{1}$ corresponds to the minimum imaging half FoV, and the ray at $\theta_{2}$ corresponds to the maximum imaging half FoV. The central region ($\theta$ < $\theta_{1}$) is the blind zone where no imaging occurs. For the LSF system demonstrated in this work, $\theta_{1}$ = 40° and $\theta_{2}$ = 95°, yielding a panoramic FoV of  $360^\circ{\times}(40^\circ{\sim}95^\circ$). 

\begin{figure*}[!h]
  \centering
  \includegraphics[width=0.7\linewidth]{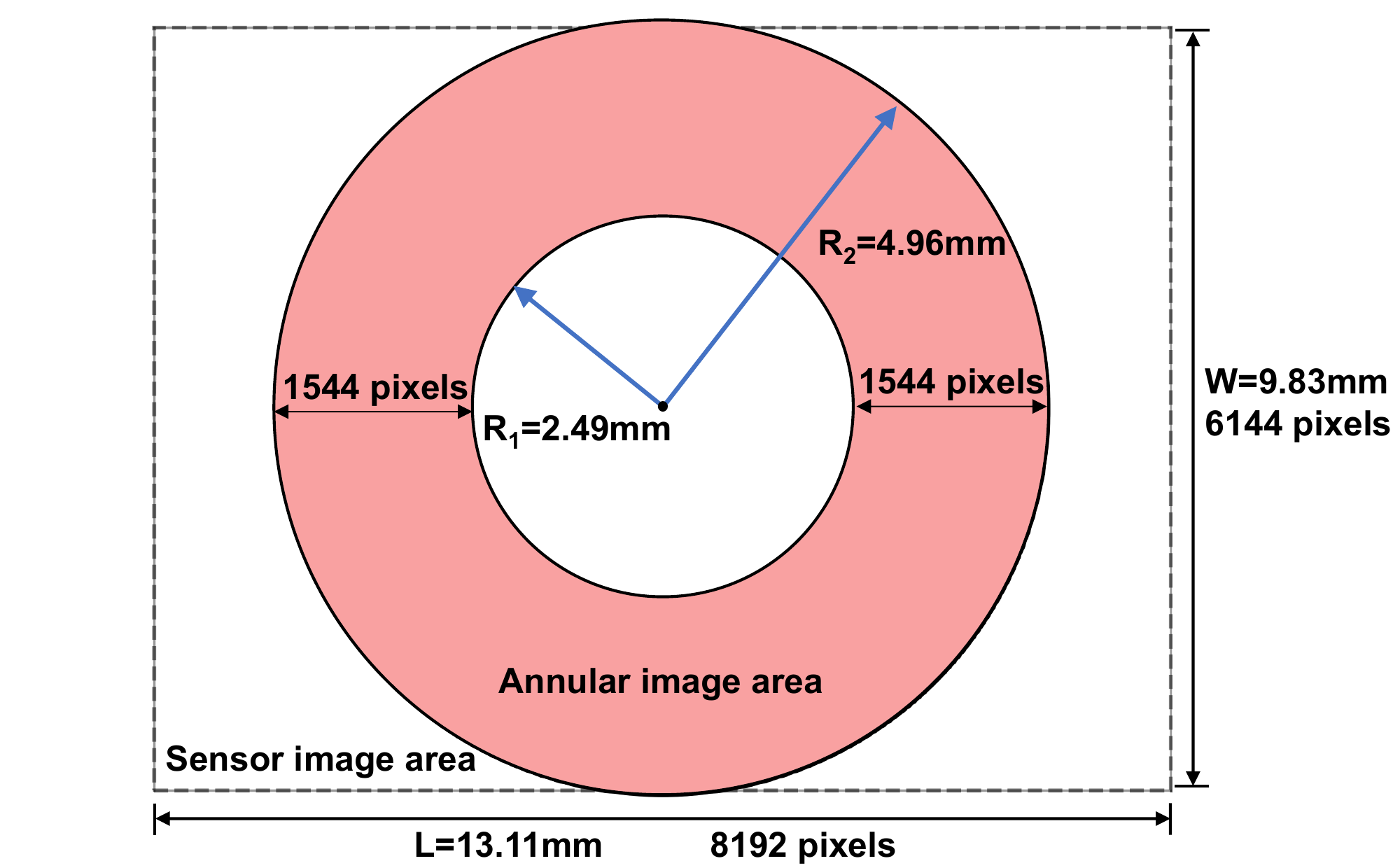}
  \caption{Schematic of the annular image area projected onto the rectangular sensor.(LSF prototype)}
  \label{fig:annular image area}
\end{figure*}

As demonstrated by the fabricated LSF prototype (this Work A in manuscript), the dimensional ratio of the MBPIS imaging area to the image sensor is illustrated in Fig.~\ref{fig:annular image area}. The minimum image height of the MBPIS is 2.49 mm, and the maximum image height is 4.96 mm. The single-side width of the original image along the optical axis is 1543 pixels, resulting in a total annular pixel occupancy of 3086 pixels across the sensor. The effective annular imaging region corresponds to approximately 22.6 megapixels.

The length of the unwrapped perspective image corresponds to the outer circumference, which is approximately 31 mm, while the inner ring is filled using bicubic interpolation. The image height is 2.47 mm. The "3K resolution" refers to the number of sensor pixels occupied horizontally across the diameter of the annular image (i.e., 3086 pixels).

This coordinate definition establishes the geometric foundation for the reverse ray tracing model detailed in Section 2.

\section{Ray tracing model of minimalist biomimetic panoramic imaging system}

As shown in Fig.~\ref{fig:sup.raytrace}, the ray tracing model of the single lens minimalist biomimetic panoramic imaging system (MBPIS). The actual ray is refracted and reflected twice by a point $P_4$ and then incident on point $P_0$ of the sensor. The incident points of each FoV correspond one-to-one with the image points on the sensor. As the design goal is to achieve the imaging principle of the far center of the image for each FoV, the imaging process of the panoramic system in Fig.~\ref{fig:sup.raytrace} can be traced back from point $P_0$ on the image plane to point $P_4$ on the object plane, which makes it more convenient to calculate the influence of various physical quantities on image height and distortion, and analyze the optimization direction of design parameters to guide.

Define the vertex of the front surface S1 as the origin $O$, the Cartesian coordinate system as the $ X-Y $ plane, the curvature radius of the front surface S1 as $r_1$, the diameter as $D_1$, the curvature radius of the rear surface S2 as $r_2$, the diameter as $D_2$, and the distance between the vertices of the front and rear surfaces as $d$. In order to facilitate the calculation of the refraction and reflection angles of light passing through the panoramic system, some auxiliary lines and angles have been added. At position $P_2$, the red dotted line is the normal of point $P_2$, and $\theta_2$ and $\theta_2'$ are the incident angle and reflection angle of the light at point $P_2$, respectively. 
At position $P_3$, the red dotted line represents the normal of point $P_3$, and the purple solid line represents the auxiliary line parallel to the optical axis passing through point $P_3$. 
The sum of $\theta_{31}$ and $\theta_{32}$ is the incident angle at point $P_3$, $\theta_{31}$ can be calculated from $\theta_2$ and $\theta_2'$, while $\theta_{32}$ can be solved by calculating the normal slope using the $P_3$ coordinate and the S2 circle equation on the back surface. $\theta_3'$ is the reflection angle, which can be obtained by solving $\theta_{31}$ and $\theta_{32}$ according to the reflection formula. 
At position $P_4$, the red dotted line represents the normal of point $P_4$, and the purple solid line represents the auxiliary line parallel to the optical axis passing through point $P_4$. $\theta_{41}$ is the incident angle, and the sum of $\theta_{41}$ and $\theta_{42}$ is equal to the sum of $\theta_{32}$ and $\theta_3'$. By calculating the normal slope based on the $P_4$ coordinate and the S1 circle equation of the front surface, $\theta_{42}$ can be obtained. $\theta_{42}$ and $\theta_{41}'$ are equal. $\theta_{42}'$ is the refractive angle of point $P_4$, which can be solved from the refractive formula. The sum of $\theta_{41}'$ and $\theta_{42}'$ is the angle relative to the optical axis, that is, the half FoV. Specifically, the light path reverse tracking path can be divided into the following 5 stages:

\begin{figure*}[!h]
  \centering
  \includegraphics[width=0.4\linewidth]{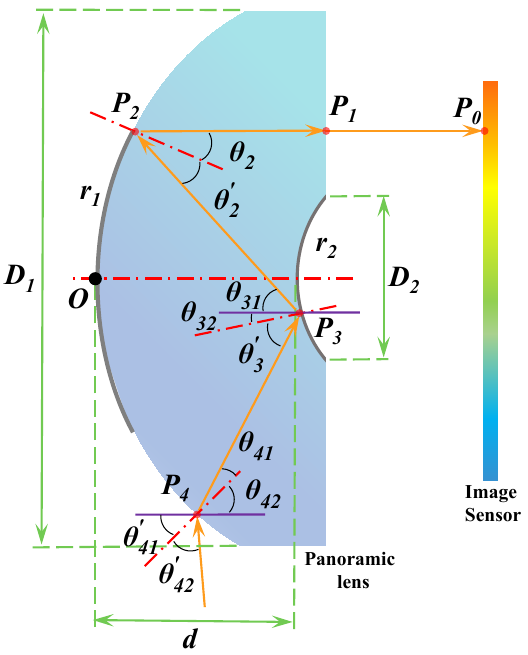}
  \caption{MBPIS ray tracing model.}
  \vskip-2ex
  \label{fig:sup.raytrace}
\end{figure*}

(1) Point $P_{0}\left(x_{0}, y_{0}\right)$ propagates to point $P_{1}\left(x_{1}, y_{1}\right)$ 
When light is transmitted from point $P_0$ to point $P_1$, for an image point $P_{0}\left(x_{0}, y_{0}\right)$ with an image height $h$, according to the design principle of image-space telecentric system, the direction of the light passing through the rear surface is parallel to the optical axis. Eq. \ref{eq:1} can be obtained.
\begin{equation}
y_{1} = y_{0} = h.
\label{eq:1}
\end{equation}

According to the circular equation of the curvature radius $r_2$ of the rear surface S2 and the diameter of the rear surface $D_2$, the coordinates $x_1$ of the edge of the rear surface can be calculated.
\begin{equation}
x_{1} = d + r_{2} - \sqrt{{r_{2}}^{2} - \left(\frac{D_{2}}{2}\right)^{2}}.
\end{equation}

(2) Point $P_{1}\left(x_{1}, y_{1}\right)$ propagates to point $P_{2}\left(x_{2}, y_{2}\right)$ 

According to the propagation process from point $P_0$ to point $P_1$, the light ray from point $P_1$ to point $P_2$ is still parallel to the optical axis. The $y$ coordinate of $P_2$ can be obtained.
\begin{equation}
y_{2} = y_{1} = h.
\end{equation}

Based on the circular equation of the front surface S1 and the coordinates $y_2$, the effective solution $x_2$ can be obtained.
\begin{equation}
x_{2} = r_{1} - \sqrt{r_{1}^{2} - h^{2}}.
\end{equation}

(3) Point $P_{2}\left(x_{2}, y_{2}\right)$ propagates to point $P_{3}\left(x_{3}, y_{3}\right)$ 

According to the circular equation of S1 on the front surface and the height coordinates of $P_2$ and the law of reflection of ray, the angle $\theta_{2}$ of the incident ray can be calculated as:
\begin{equation}
\theta_{2} = \arcsin\left(\frac{y_{2}}{r_{1}}\right).
\end{equation}

The angle $\theta_{2}$ of the reflected light is:
\begin{equation}
\theta_{2} = \theta_{2}^{\prime}.
\end{equation}

For a straight line $P_2$$P_3$, according to the angle of incident and reflected ray, the linear slope $k_{P_2P_3}$ and linear equation can be obtained. It can be associated with the S2 circular equation of the rear surface as a system of equations as follows.
\begin{equation}
k_{P_2P_3} = -\tan(\theta_2 + \theta_2').
\end{equation}
Solve the following set of equations to obtain the coordinates of $P_3$.
\begin{equation}
\begin{cases}
y_3 - y_2 = -\tan(\theta_2 + \theta_2') \cdot (x_3 - x_2), \\
\left[x_3 - (d_1 + r_2)\right]^2 + y_3^2 = r_2^2.
\end{cases}
\end{equation}

(4) Point $P_{3}\left(x_{3}, y_{3}\right)$ propagates to point $P_{4}\left(x_{4}, y_{4}\right)$ 

According to the geometric relationship, the incidence angle $\theta_{3}$ can be divided into two parts:
\begin{equation}
\theta_{3} = \theta_{31} + \theta_{32}.
\end{equation}
According to the geometric relationship of ray, $\theta_{31}$ can be obtained. 
\begin{equation}
\theta_{31} = \theta_{2} + \theta_{2}^{\prime}.
\end{equation}

According to the law of refraction, $\theta_{32}$ can be calculated.
\begin{equation}
\theta_{32} = \arcsin\left(\frac{y_3}{r_2}\right).
\end{equation}

Since the reflection angle $(\theta_{3}^{\prime}$ has the same value as the incidence angle $\theta_{3}$. For the line $P_3$$P_4$, the slope $k_{P_3P_4}$ can be computed as:
\begin{equation}
k_{P_3P_4} = \tan(\theta_{3}^{\prime} + \theta_{32}).
\end{equation}

Associate the equation of the line $P_3$$P_4$ with the circular equation on the front surface S1; the equation can be expressed as follows. 
Then, the coordinates of $P_4$ can be obtained.
\begin{equation}
\begin{cases}
y_4 - y_3 = \tan(\theta_3' + \theta_{32}) \cdot (x_4 - x_3), \\
(x_4 - r_1)^2 + y_4^2 = r_1^2.
\end{cases}
\end{equation}

(5) Point $P_{4}\left(x_{4}, y_{4}\right)$ propagates to the incident FoV of the object space
\begin{equation}
\theta_{4} = \theta_{41} + \theta_{42} = \theta_{3}' + \theta_{32}.
\end{equation}
According to the geometric relationship and the law of refraction, $\theta_{41}'$ and $\sin \theta_{42}'$ can be obtained.
Solve the above equation for the coordinates.
\begin{equation}
\theta_{41}' = \theta_{42},
\end{equation}
\begin{equation}
\sin \theta_{42}' = n \cdot \sin \theta_{41}.
\end{equation}

The angle of the incident angle $\theta_{\text{input}}$ is the sum of the incident angle $\theta_{41}'$ and refraction angle $\theta_{42}'$.
\begin{equation}
\theta_{\text{input}} = \theta_{41}' + \theta_{42}'.
\end{equation}

It is known that the distortion formula is the following formula, where the image plane height is $h$, the focal length of the system is $f$, and the incident angle is $\theta$.
\begin{equation}
D_{f-\theta} = 100\% \times \frac{h - f \cdot \theta_{\text{input}}}{f \cdot \theta_{\text{input}}}.
\end{equation}

According to the above calculation procedure, the $f-\theta$ distortion can be calculated at different fields of view.
According to the image height and its corresponding incident FoVs in the above calculation process, the average angular resolution within the image plane is calculated at intervals of 0.1 mm, which is plotted as concentric rings of different colors and marked with the corresponding angular resolution values of different colors.

The design parameters of the final seven different minimalist biomimetic computational panoramic imaging systems are shown in Table~\ref{tab:1}. The first five systems are designed using standard spherical surfaces, while ASF and LASF are panoramic imaging systems designed using even aspheric surfaces.

\begin{table}[!t]
\centering
\scriptsize
\caption{\bf Design Parameters of Biomimetic Computational Panoramic Imaging Systems}
\setlength{\tabcolsep}{3pt}
\begin{tabular}{cccccccc}
\hline
System & SF & LSF & SSC & LSSC & SNC & ASF & LASF \\
\hline
RS1 & 15.86 & 16.01 & 16.17 & 16.31 & 11.68 & 15.35 & 13.73\\
RS2 & 7.17 & 6.17 & 6.34 & 5.39 & 8.79 & 16.65 & 17.78\\
CT & 10.02 & 10.08 & 9.91 & 9.98 & 6.43 & 10.10 & 8.68\\
DVA & 0.21 & 0.41 & 0.30 & 0.64 & 5.11 & 0.47 & 0.52\\
MI & 2.00 & 2.00 & 2.00 & 2.00 & 2.00 & 1.46 & 1.46\\
MA & 25.43 & 25.43 & 25.43 & 25.43 & 25.43 & 90.27 & 90.27\\
S1K & - & - & - & - & - & -1.07 & -2.60\\
S1A4 & - & - & - & - & - & 4.412E-05 & 1.345E-04\\
S1A6 & - & - & - & - & - & 3.247E-07 & 3.118E-07\\
S1A8 & - & - & - & - & - & -5.575E-09 & -8.880E-09\\
S2K & - & - & - & - & - & -0.68 & 1.57\\
S2A4 & - & - & - & - & - & 4.081E-05 & 1.075E-04\\
S2A6 & - & - & - & - & - & -1.419E-05 & -2.002E-05\\
S2A8 & - & - & - & - & - & 7.374E-08 & 3.431E-07\\
\hline
\end{tabular}
  \label{tab:1}
\end{table}

\subsection{Optical micro-channel glass array sensor}

\begin{figure*}[!h]
  \centering
  \includegraphics[width=0.4\linewidth]{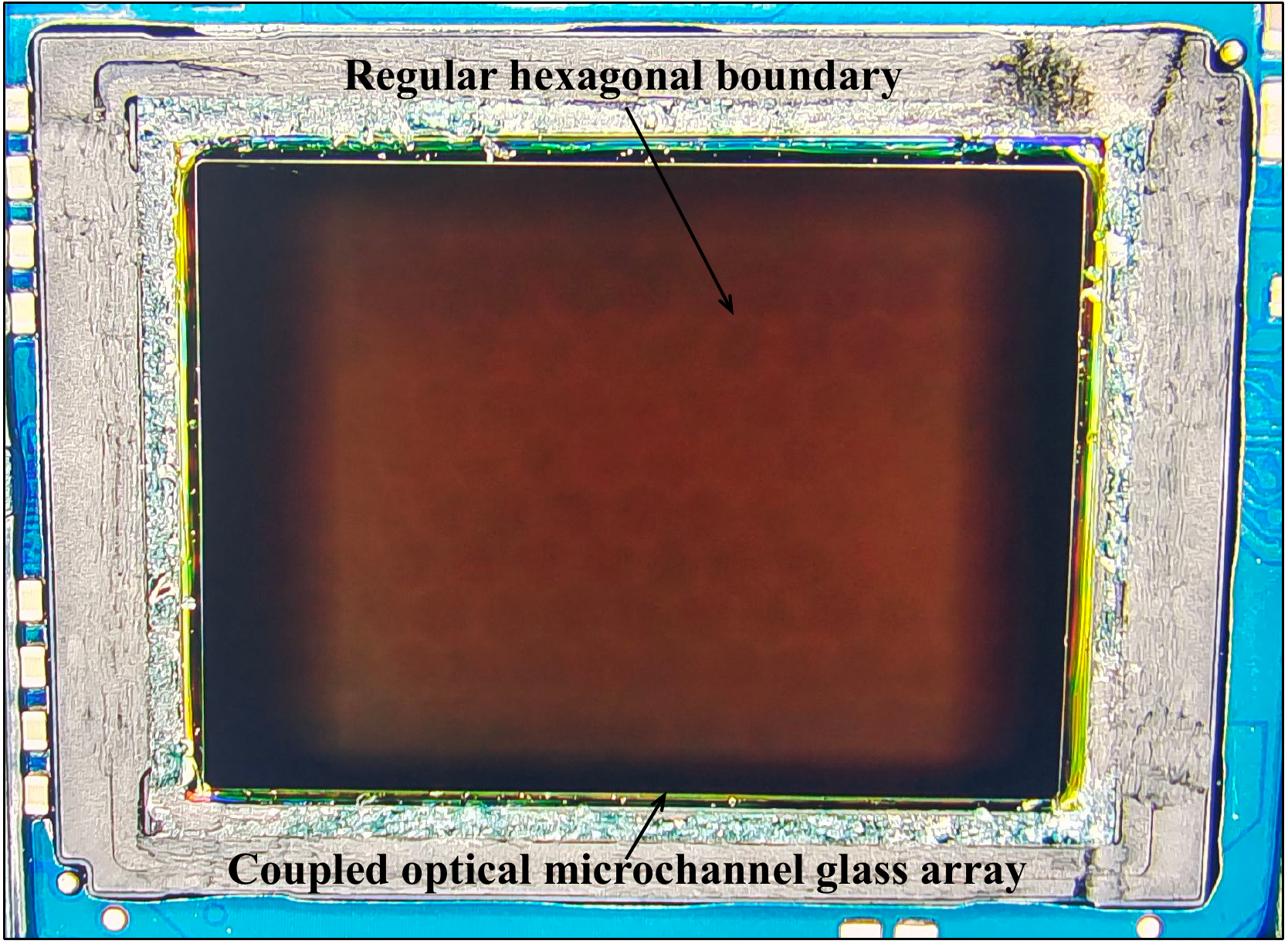}
  \caption{Coupling detection of a new type of optical micro-channel glass array sensor.}
  \vskip-2ex
  \label{fig:Coupled Optical microchannel glass array sensor}
\end{figure*}

We use the smartphone image sensor Sony IMX 989 as the sensor to be coupled, and customize the optical micro-channel array to the effective pixel area size of the sensor. It is worth noting that due to the strict alignment of the lateral and longitudinal dimensions of the optical micro-channel array during the coupling process, microscopic coupling detection of the optical micro-channel glass array sensor is required during coupling, as shown in Fig. \ref{fig:Coupled Optical microchannel glass array sensor}. 
At the same time, in order to avoid the occurrence of Moire fringes during imaging between the optical micro-channel glass array and the micro-channel glass array arranged in a regular manner in the mobile phone image sensor, attention should also be paid to the image on the display screen during coupling to achieve the minimum Moire fringes in the coupled image.

\begin{figure*}[!h]
  \centering
  \includegraphics[width=0.4\linewidth]{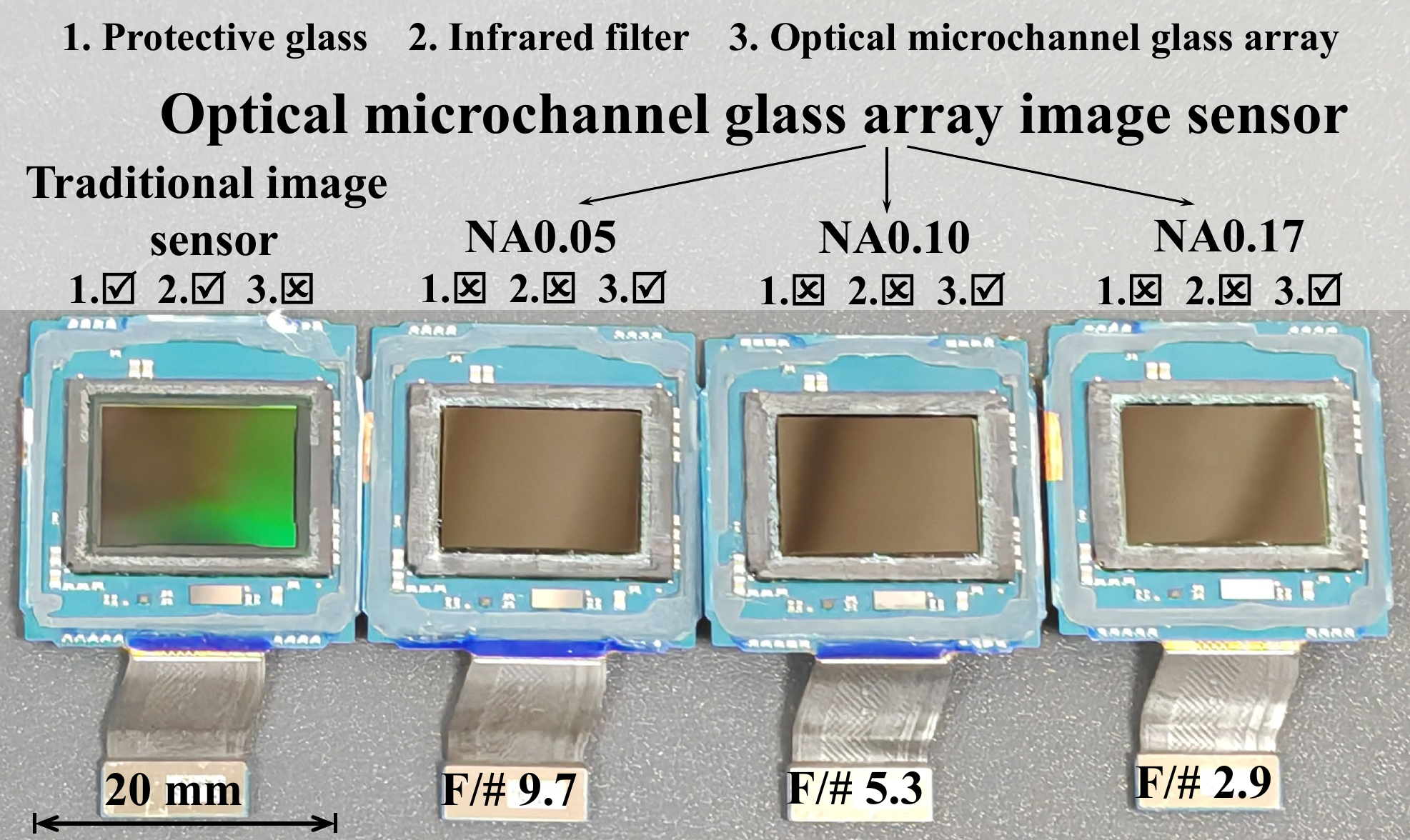}
  \caption{A novel image sensor obtained by coupling traditional image sensors with optical micro-channel glass arrays.}
  \vskip-2ex
  \label{fig:Optical microchannel glass array sensor}
\end{figure*}

As shown in Fig.~\ref{fig:Optical microchannel glass array sensor}, we constructed new prototypes of optical micro-channel glass sensors with NA values of 0.05, 0.10, and 0.17, respectively. Due to the coupling of glass-based optical micro-channel glass sensors, the protective glass used in traditional image sensors is no longer needed in this new type of image sensor. Meanwhile, due to the fact that the final refractive transmission surface of our designed single lens panoramic system is flat, an infrared cutoff film can be coated on this flat surface to achieve visible light imaging. Therefore, this new image sensor does not require additional infrared filters and can further reduce the hardware thickness by about 1 mm in the direction perpendicular to the optical axis of the image sensor, thus achieving the compact advantage of a single-lens panoramic system. In order to reduce the errors caused by processing, we selected an LSF system with a larger FoV for processing. This system has the highest aberration among the 7 designs, so using it for imaging calculations is representative. The optical micro-channel glass array with an NA of 0.10 corresponds to an $F$$/$\# of 5.3 in the LSF system. In order to demonstrate the different F-numbers of imaging results with different thicknesses, we simultaneously combined optical micro-channel glass array sensors with NA values of 0.05 and 0.17 with LSF lenses for imaging. Based on the incident angle, transmittance, and LSF design results of the square ray, the system $F$$/$\# of the optical micro-channel glass array sensor combined with NA values of 0.05 and 0.17 can be obtained as 9.7 and 2.9, respectively. 
If a larger or smaller aperture is required, the control of the system $F$$/$\# can also be achieved by further reducing or increasing the thickness of the optical micro-channel glass array.

\begin{figure*}[!h]
  \centering
  \includegraphics[width=0.78\linewidth]{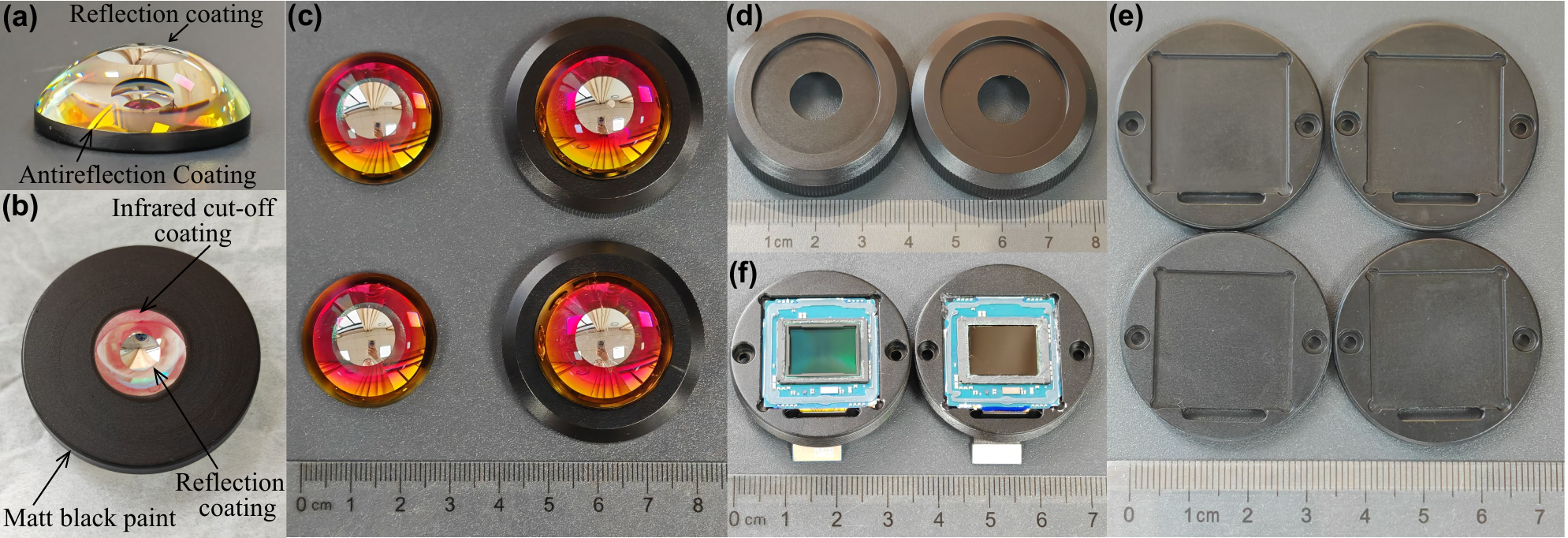}
  \caption{Prototype components of minimalist biomimetic computational panoramic imaging system. (a) Side view of spherical minimalist biomimetic computational panoramic imaging system based on LSF lens (b) Bottom view of spherical minimalist biomimetic computational panoramic imaging system based on LSF lens (c) System size with only lens and lens mount installed (d) Lens mount (e) Chip base with traditional image sensor and new optical micro-channel glass array installed (f) Internal structure of chip base.}
  \vskip-2ex
  \label{fig:System components}
\end{figure*}

The spherical minimalist biomimetic panoramic imaging system based on the LSF lens is shown in Fig.~\ref{fig:System components}(a). Its first refractive surface is coated with an anti-reflective film, which can improve the transmittance of the imaging beam and reduce reflection and stray light. The center of the first refractive surface is the second reflective surface, coated with a reflective film. 
The bottom view of the single lens system is shown in Fig.~\ref{fig:System components}(b), with an infrared cutoff film coated on the second refractive surface and a reflective film coated on the first reflective surface in the center. The average transmittance of the infrared cutoff film is greater than 95\% in the wavelength range of 420-630 nm, and the minimum transmittance is greater than 92\%. When the wavelength is between 640-660 nm, the transmittance of the wavelength is about 50\%. When the wavelength is between 700-1000 nm, the maximum transmittance is less than 1\%, and the average transmittance is less than 0.5\%. Due to the high spectral response of mobile phone image sensors to only visible and near-infrared light bands, light in other bands will not have a significant impact on imaging on the image sensor. Due to the high reflectivity of the second refractive surface coated with an infrared cutoff film for red and near-infrared orange spectral light, it can be observed from  Fig.~\ref{fig:System components}(c) that in the top view of the lens, a ruby-like luster can be observed, while the central reflective film presents a silver luster. Compared to just the lens, the lens size is larger with the addition of a lens mount. The structural design of the lens mount is shown in  Fig.~\ref{fig:System components}(d). 
The function of this lens mount is to screw with the chip base in  Fig.~\ref{fig:System components}(e), thereby achieving the rear focus function of the lens. The lens mount and chip base in this prototype are made of metal and can also be designed with adhesive bonding using plastic injection molding technology. They can be modified to other smaller mechanical structures to further reduce system weight and volume, making them suitable for mass production. The chip base can fix image sensors and new optical micro-channel glass array sensors, making it convenient for optical systems to be installed on various lightweight mobile intelligent computing platforms.

\begin{figure*}[!h]
  \centering
  \includegraphics[width=0.78\linewidth]{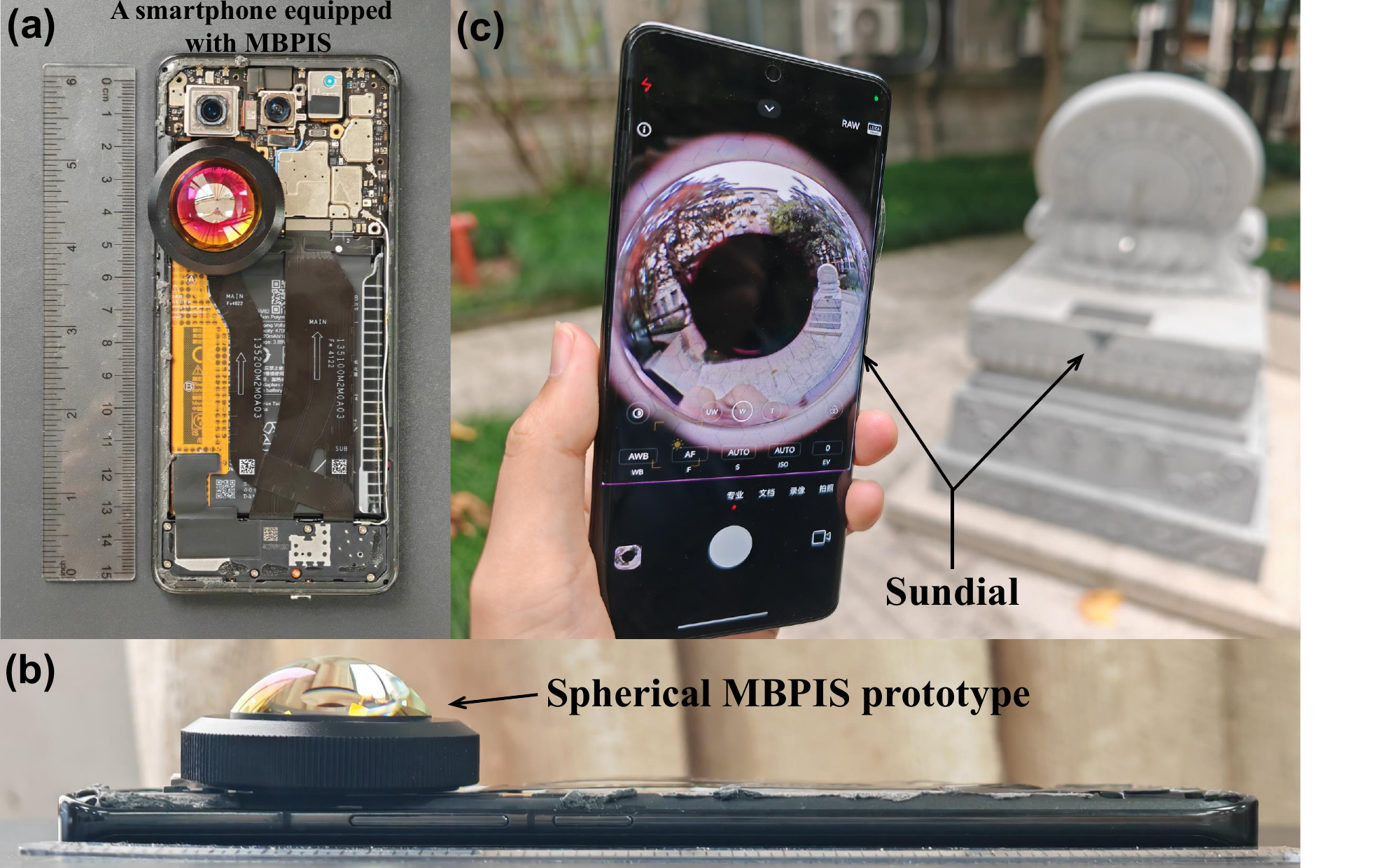}
  \caption{Prototype of a smartphone imaging system with MBPIS installed. (a) Top view. (b) Side view. (c) Compared to the image field obtained by the main camera of a mobile phone with 1$\times$ magnification, smartphones equipped with MBPIS can obtain a larger panoramic imaging FoV.}
  \vskip-2ex
  \label{fig:Prototype imaging results}
\end{figure*}

To facilitate the verification of the panoramic perception capability of the minimalist biomimetic computational panoramic imaging system on miniaturized intelligent and lightweight computing perception terminal devices. We used the IMX989 smartphone sensor with a 1-inch imaging area image sensor with a large target surface and imaged using the image sensor with protective glass and infrared filter, as well as the new optical channel glass array sensor with three different specifications of NA shown in Fig.~\ref{fig:Optical microchannel glass array sensor}. The smartphone equipped with a minimalist biomimetic computational panoramic imaging system is shown in Fig.~\ref{fig:Prototype imaging results}(a), and the side view is shown in Fig.~\ref{fig:Prototype imaging results}(b). The entire image shown in Fig.~\ref{fig:Prototype imaging results}(c) is captured using a 1x magnification mobile phone lens as the main camera, which simultaneously captures a smartphone equipped with a minimalist biomimetic computational panoramic imaging system and a sundial. However, a 1x magnification mobile phone lens can only capture a very small area of FoV except for the sundial, while a smartphone equipped with a minimalist biomimetic computational panoramic imaging system can capture a super hemisphere imaging FoV, which can obtain more surrounding environmental imaging information in addition to the sundial.

\begin{figure*}[!h]
  \centering
  \includegraphics[width=0.7\linewidth]{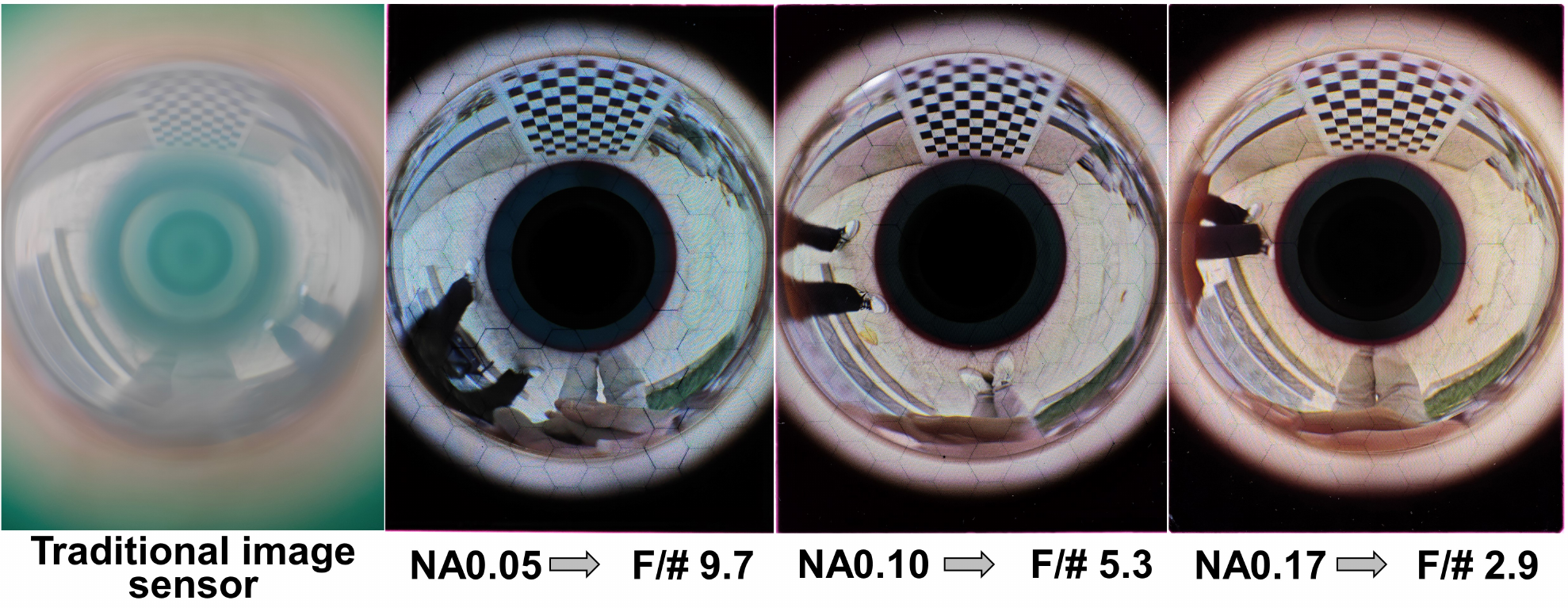}
  \caption{Panoramic imaging results of traditional image sensors and microchannel sensors with different thicknesses.}
  \vskip-2ex
  \label{fig:suppFactimg}
\end{figure*}

In the actual imaging process, we use traditional image sensors with protective glass and infrared filters as shown in Fig.~\ref{fig:Optical microchannel glass array sensor}, as well as three specifications of NA microchannel array image sensors, combined with MBPIS based on the LSF system for imaging. The actual imaging results are shown in Fig.~\ref{fig:suppFactimg}.

When using traditional mobile phone image sensors for imaging, the image is blurry due to the inability to filter out stray light outside the imaging beam. Additionally, due to the thickness gap in front of the image caused by the protective glass and infrared filter, the imaging plane of the LSF system with a shorter back focus cannot coincide with the plane where the sensor pixels are located, resulting in a smaller imaging circle. Due to multiple reflections of the imaging beam between the protective glass and the infrared filter, concentric circular ghost images appear in the blind zone. After using the microchannel array image sensor with different NA in this paper, there is no need to protect the glass and infrared filter, achieving absorption of non imaging beams and avoiding the entry of stray light, resulting in clearer imaging images.
When the aperture is larger and the $F$$/$\# is smaller, the amount of incoming light is greater, and the imaging results are brighter. 
Due to the increased aperture related aberrations (spherical aberration, positional chromatic aberration) caused by the small $F$$/$\# and large relative aperture, the imaging quality is more blurry. 
Therefore, the $F$$/$\# of the panoramic system can be controlled by manufacturing microchannel arrays with different NA to meet the aperture requirements of the panoramic system in different application scenarios. The control experiment of $F$$/$\# mentioned above is only to verify the constraint ability of the microchannel sensor on light, and no targeted aberration optimization design has been made for the system based on $F$$/$\#. Therefore, in the later calculation of the imaging process, the LSF system with $F$$/$\# 5.3 will still be used for the experiment.

\section{Data construction pipeline}

For training a learning-based model to enhance the imaging quality of MBPIS, data pairs of degraded MBPIS images and their aligned clear images (the ground truth) are significant.
To this end, our data construction pipeline starts from high-quality panoramic annular images, simulating all potential image degradations of MBPIS on them to produce MBPIS's imaging results.
Specifically, $3$ main degradation factors are considered: the image distortion induced by reduced FoV, the residual optical aberrations induced by the lens design, and the fix-pattern noise induced by the micro-channel glass array sensor.

\subsection{FoV adaptation}
\begin{figure*}[!h]
  \centering
  \includegraphics[width=0.75\linewidth]{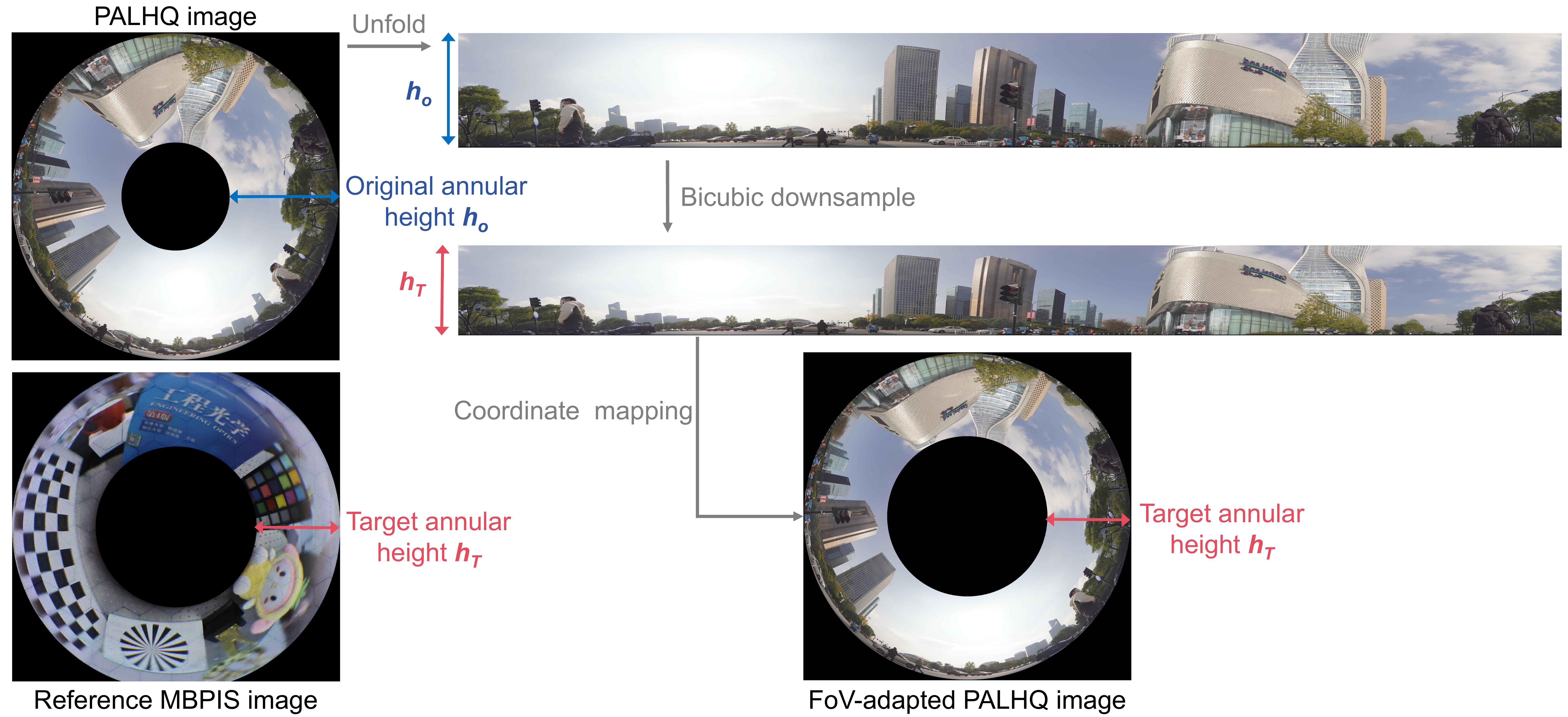}
  \caption{\textbf{Illustration of the FoV adaptation process.}}
  \vskip-2ex
  \label{fig:fova}
\end{figure*}

FoV adaptation is applied to make sure that the blind area of PALHQ images is equivalent to that of MBPIS images, which also simulates additional image distortion due to reduced FoV (the FoV of MBPIS is smaller than that of the panoramic lens in PALHQ, leading to a larger blind area).
To be specific, a real-world MBPIS image is employed as a reference, from which the pixel radius of the blind area and annular band thickness are extracted as target dimensions. 
Subsequently, the PALHQ image undergoes bicubic interpolation to match its resolution precisely with that of the MBPIS image.
Following this, the image is unfolded via coordinate mapping, producing a perspective image where the height corresponds to the thickness of the annular band, and the width equals the circumference of the outer ring. 
Any missing inner ring pixels are filled in using bicubic interpolation. 
The unfolded PALHQ image is then down-sampled in height via bicubic interpolation to match the annular band thickness of the reference MBPIS, thereby yielding the FoV-adapted unfolded image. 
Finally, coordinate mapping is applied to project the unfolded image back onto the original annular image, producing the FoV-adapted high-quality panoramic annular image.
The schematic diagram of the FoV adaptation process is illustrated in Fig.~\ref{fig:fova}.

\subsection{Aberration simulation}

\begin{figure*}[!h]
  \centering
  \includegraphics[width=0.7\linewidth]{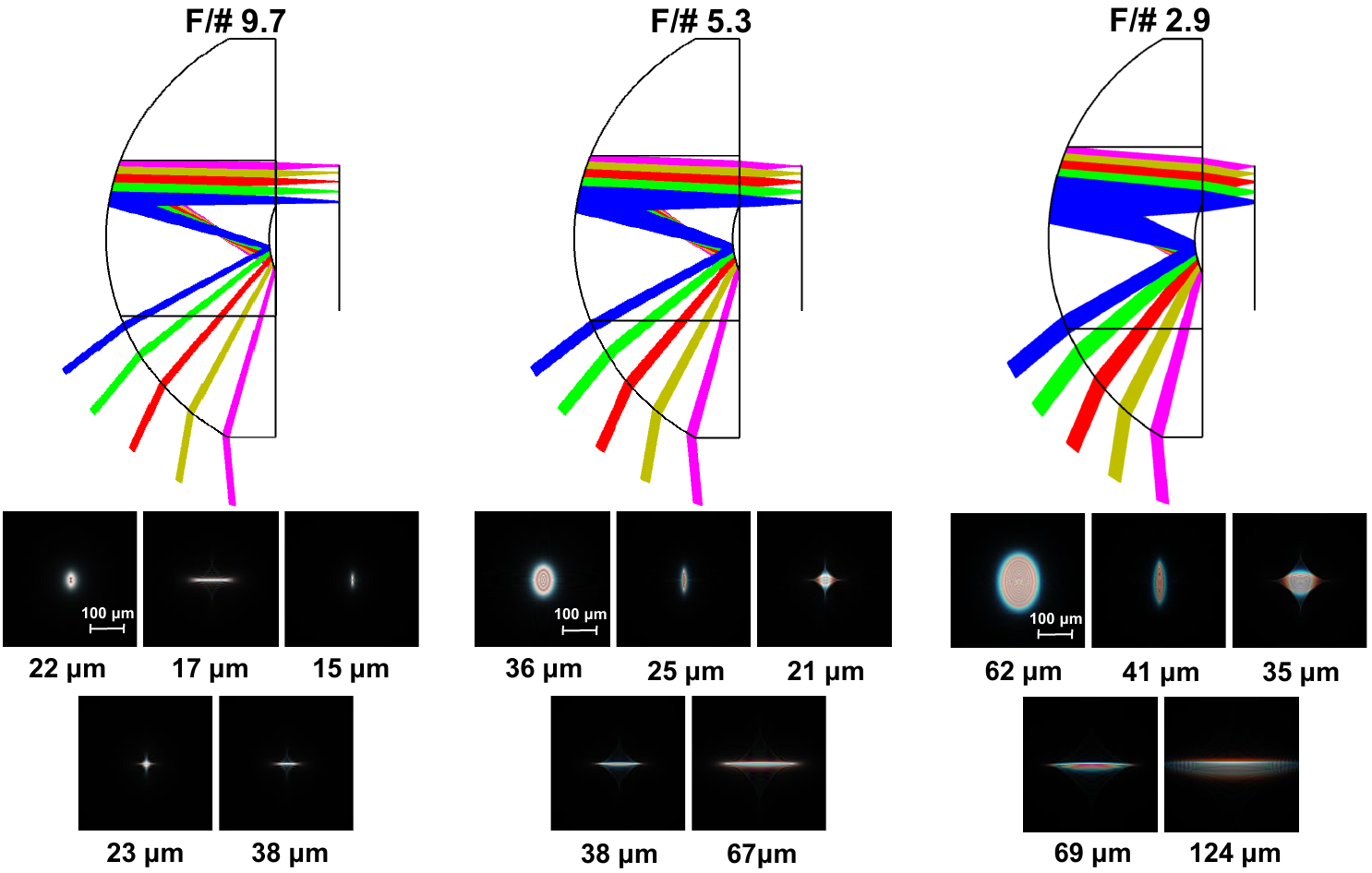}
  \caption{\textbf{Lens layouts for $3$ F-numbers and the corresponding PSF for $5$ samples with RMS spot radii.} The side length of each PSF image is $0.4096$ mm.}
  \vskip-2ex
  \label{fig:psf}
\end{figure*}

\begin{figure*}[!h]
  \centering
  \includegraphics[width=0.6\linewidth]{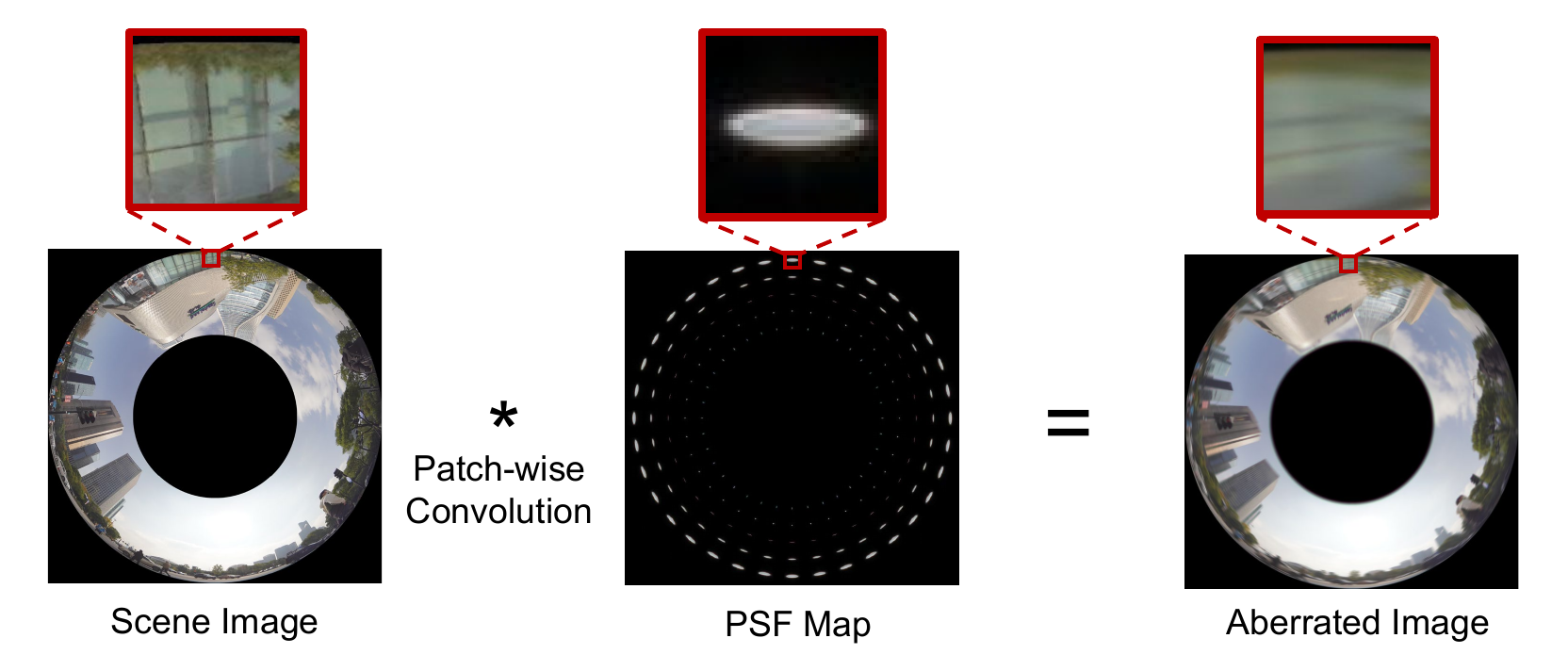}
  \caption{\textbf{Degradation model.}  The degradation caused by aberrations is characterized by PSF. 
  Our simulation model applies the patch-wise convolution between the scene image and PSF map for generating the aberrated image.}
  \vskip-2ex
  \label{fig:conv}
\end{figure*}

In our imaging simulation pipeline, the degradation caused by aberrations is characterized by the energy dispersion of the Point Spread Function (PSF). 
We obtain the accurate PSF from Zemax. 
Specifically, we use Zemax to compute the diffraction PSF using direct integration of the Huygens wavelets method. 
Fig.~\ref{fig:psf} shows the lens layouts for $3$ F-numbers and the corresponding PSFs for $5$ samples with the LSF system.
The PSF dispersion degree calculated by the simulation design in Fig.~\ref{fig:psf} is consistent with the actual imaging experiment in Fig.~\ref{fig:suppFactimg}, verifying the constraint effect of the microchannel array on the NA of the imaging beam in the image direction. Due to the fact that the LSF system has a spherical optical surface type and a planar microchannel end face in the design results of this paper, we will still use the PSF with the original design F-number of 5.3 for actual aberration simulation in the future.
Our simulation model applies patch-wise convolution between the FoV-adapted PALHQ image $I_S$ and the PSF map to generate the aberrated image $I_A$. 
Initially, the FoV-adapted PALHQ image $I_S$ is divided into $n_h{\times}n_w$ patches. Each patch is designated as $I_S(c, h, w)$ with $s{\times}s$ pixels, where $c$ indicates one of the R, G, and B channels, and $h$ and $w$ denote the position of the patch on the image plane. The associated $PSF (c, h, w)$ is derived by interpolating PSFs across all sampled fields of view and adjusting them by rotating to the correct angle:
\begin{equation}
\label{eq:simu4}
    PSF(c, h, w) = P_{rot}(\sum_{\theta} W(\theta)\cdot PSF(c, \theta)),
\end{equation}
where $PSF(c, \theta)$ is the PSF from a specific FoV and $W(\theta)$ is the normalized interpolation weight determined by the inverse square formula. In addition, $P_{rot}$ represents the rotation process, and the rotation angle $\beta$ can be calculated by $\beta = \arctan(w/h)$, which corresponds to rotating the PSF in the $x$-direction to align with the sensor position $(h, w)$. The degraded image patch $I_A(c, h, w)$ is approximated as:
\begin{equation}
\label{eq:simu5}
    I_A(c, h, w) \approx PSF(c, h, w) \ast I_S(c, h, w).
\end{equation}
After that, we stitch $n_h{\times}n_w$ degraded image patches to obtain the complete aberration image $I_A$ in the same order as we previously split $I_S$. Fig.~\ref{fig:conv} illustrates the overall process of patch-wise convolution.

\subsection{Fix-pattern noise simulation}
\begin{figure*}[!h]
  \centering
  \includegraphics[width=0.7\linewidth]{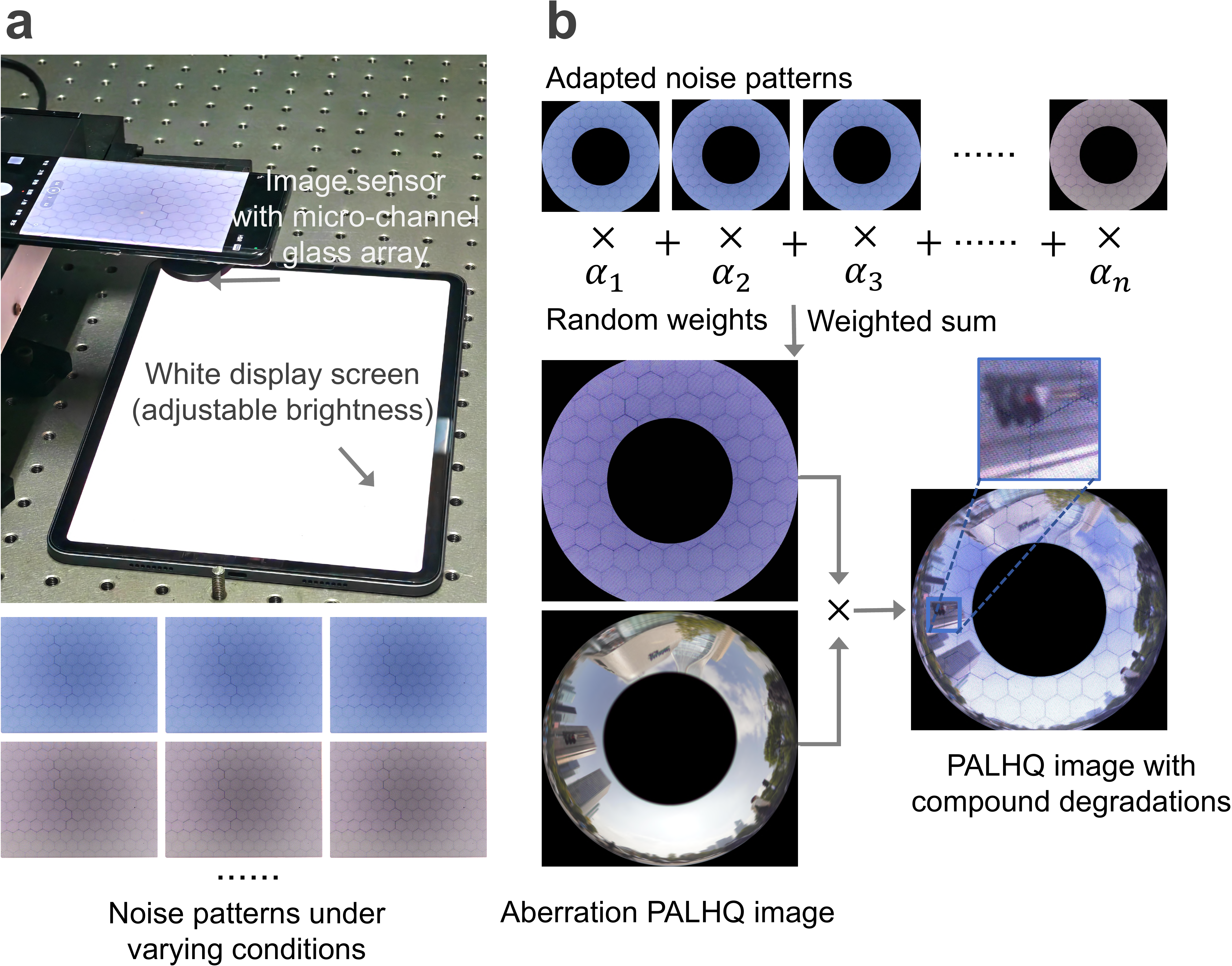}
  \caption{\textbf{Pipeline for fix-pattern noise calibration and simulation.} \textbf{a} Experimental setup of calibration and the calibrated noise patterns under varying conditions by adjusting the display brightness, shooting distance, and imaging angle. \textbf{\textbf{b}} We use random weights to generate weighted summed noise pattern to simulate the fix-pattern noise on the PALHQ aberration image by multiplying, which can simulate the realistic noise patterns of micro-channel glass array.}
  \vskip-2ex
  \label{fig:noise}
\end{figure*}

Fix-pattern noise constitutes the other significant source of image degradation of MBPIS images. 
Consequently, it is imperative to calibrate the precise noise pattern and apply it to the aforementioned simulated aberration images for end-to-end correction of these compound degradations based on supervised training.
As depicted in Fig.~\ref{fig:noise}a, we employ a sensor equipped with the micro-channel glass array but devoid of a lens to directly image a white display screen for the calibration of fix-pattern noise. 
This calibration experiment is conducted in a low-light laboratory setting. Furthermore, to account for the complexities of real-world imaging scenarios, we acquire multiple sets of noise patterns under varying conditions by adjusting the display brightness, shooting distance, and imaging angle.
Then, as shown in Fig.~\ref{fig:noise}b, we adopt the concept of data augmentation to introduce a plethora of random combinations of these multiple noise patterns into the training data, thereby endowing the model with robust processing capabilities for such fixed-pattern noise. 
Specifically, for each aberration image intended for noise simulation, we generate a set of random weights, which are applied to all noise patterns to produce a weighted sum, thus creating a random noise pattern to serve as the noise source for this image. 
Additionally, empirical observations reveal that this fix-pattern noise manifests as dark textures (gray value of $0$) in white images (gray value of $255$), indicating its multiplicative nature. 
Consequently, we simulate this noise by multiplying the noise pattern with the aberration image.
It is noteworthy that, although the multiplicative noise is derived from our hypothesis, the remarkably close resemblance between the simulated and reference noise images, coupled with the subsequent successful denoising results, substantiates the validity of this conjecture.

\subsection{Adaptation for ground-truth}
We observe that the aforementioned noise simulation introduces additional brightness modulation and color bias to the images. To prevent the model from learning unnecessary variations in brightness and color, we apply corresponding adjustments to the ground truth, modulating its brightness and color accordingly. This step constitutes the ``adaptation for ground-truth'' stage of our simulation pipeline.
In practice, we first calculate the mean grayscale values of the RGB channels for both the degraded PALHQ image and the ground truth image. For each color channel of the ground-truth image, we normalize it using its corresponding mean grayscale value and then multiply it by the mean grayscale value of the same channel in the degraded image, thereby achieving the desired brightness and color modulation. 
Ultimately, the adapted ground truth and its corresponding degraded PALHQ image form the training data pair.

\section{Implementation of PIENet}
The code of PIENet will be made publicly available at \url{https://github.com/zju-jiangqi/PIENet} to make it convenient for future researchers to replicate and understand our work.

\subsection{Network architecture}
The image enhancement network $F_{E}$ can be trained on $\{I_{ab}, I_{gt}\}$ via supervised learning. 
Nevertheless, this task is an ill-posed problem, where the details of the high-quality image corrupted by the compound degradation are hard to retrieve.
Without any useful image prior, the network can hardly recover the lost textures and details.

Inspired by the pioneering works on image generation models~\cite{rombach2022high,zhang2023regularized}, we develop a prior-embedded image enhancement model (PIENet) that introduces high-quality priors (HQP) to mitigate the issue.
The core idea of PIENet is to learn the distribution of clear, high-quality images, which are quantized to guide the image enhancement process.
We implement this process using the Vector-Quantized (VQ) codebook learning in VQGAN~\cite{van2017neural,esser2021taming}, which requires a smaller training scale compared to diffusion models~\cite{ho2020denoising} without introducing significant additional computational overhead during inference.

In the VQGAN pre-training stage~\cite{chen2022real}, the clear images are encoded into a discrete latent space to learn a codebook characterizing the high-quality image features.
Given a codebook $\mathcal{Z}=\{z_k\}_{k=1}^{K}\in\mathbb{R}^{n}$, the latent feature  $\hat{z}\in\mathbb{R}^{h{\times}w{\times}n}$ of the input image with spatial size $(h\times{w})$ and channel dimension $n$, is quantized by finding the nearest neighbours in $\mathcal{Z}$ for its each element $\hat{z}_{ij}$, to calculate the discrete representation ${z}^{q}\in\mathbb{R}^{h{\times}w{\times}n}$:
\begin{equation}
\label{eq:codematching1}
{z}^{q}_{ij} = \arg\min\limits_{z_k\in\mathcal{Z}}(\Vert{\hat{z}_{ij}-z_k}\Vert_2),
\end{equation}
where $K$ denotes the codebook size and $i\in\{1,2,\cdots,h\}$, $j\in\{1,2,\cdots,w\}$ denote the coordinates in the feature space. 
The quantized feature ${z}^{q}$ is then applied to reconstruct the input image for self-supervised learning.
The VQGAN pre-training stage follows the following training objective~\cite{chen2022real,esser2021taming}:

\begin{multline}
\label{eq:vqloss}
\mathcal{L}_{VQGAN} = {\Vert{F_{VQGAN}(I_{gt})-I_{gt}}\Vert_1}+ \lambda_{per}\mathcal{L}_{per}  \\
+ \lambda_{adv}\mathcal{L}_{adv} + \lambda_{codebook}\mathcal{L}_{codebook},
\end{multline}
where $F_{VQGAN}$ is the VQGAN model, and $\{\lambda_{per}, \lambda_{adv}, \\ \lambda_{codebook}\}$ is a set of hyperparameters for balancing the pre-training process. We choose LPIPS loss for $\mathcal{L}_{per}$, hinge loss for $\mathcal{L}_{adv}$, while following FeMaSR and VQGAN~\cite{chen2022real} to use $\mathcal{L}_{codebook}$ for regularizing the quantized features.
After training VQGAN, the codebook represents the priors for the input image domain, which delivers potential in guiding the downstream image restoration tasks~\cite{gu2022vqfr,zhou2022towards}. 

During the image enhancement process, encoded deep features of an MBPIS image are progressively mapped into such a latent space with an enhancement module of convolution or Transformer blocks in low-level vision~\cite{wang2018esrgan,liang2021swinir,zamir2022restormer,zhou2023srformer}, i.e., the Residual Swin-Transformer Block (RSTB)~\cite{liang2021swinir} in our case.
Then, the enhanced features, denoted by $\hat{f}$, serve two purposes: (i) reconstructing the high-quality clear image by an HQP-guided decoder group $Dec$, and (ii) generating HQP via feature matching with the codebook to guide the reconstruction process.
Specifically, for $\hat{f}$ with spatial size $(h\times{w})$ and channel dimension $n$, similar to the process in VQGAN, the nearest neighbours are found in the learned codebook $\mathcal{Z}$ for its each element $\hat{f}_{ij}$, to obtain the quantized features $\hat{f}^{q}\in\mathbb{R}^{h{\times}w{\times}n}$:
\begin{equation}
\label{eq:codematching}
\hat{f}^{q}_{ij} = \arg\min\limits_{z_k\in\mathcal{Z}}(\Vert{\hat{f}_{ij}-z_k}\Vert_2).
\end{equation} 
Furthermore, the pre-trained decoder group $Dec_{vq}$ in VQGAN is employed to produce HQP features from $\hat{f}^q$, which are used to guide the reconstruction features in $Dec$ through a fusion module $Fusion$: 
\begin{equation}
\label{eq:fusion}
\hat{f}'_{i+1} = Dec(Fusion(\hat{f}'_i, \hat{f}'^q_i), i+1), \hat{f}'^q_i = Dec_{vq}(\hat{f}'^q_{i-1}, i),
\end{equation}
where the features $\hat{f}'_i$ and $\hat{f}'^q_i$ from the $i$th layer of $Dec$ and $Dec_{vq}$ are fused by $Fusion$, and then fed to the $(i+1)$th layer of $Dec$ to produce reconstructed features $\hat{f}'_{i+1}$.
In PIENet, we take the HQP features as an additional modality, so that the fusion module can be any common multi-modal fusion method, e.g., add, concatenation (Concat.), Deformable Convolution Network (DCN)~\cite{zhu2019deformable}, Multi-modal Cross Attention (MCA)~\cite{sun2022event}, and Spatial Feature Transformation (SFT)~\cite{wang2018recovering}. 
In our case, we adopt SFT for its superior performance.
Referring to FeMaSR~\cite{chen2022real}, the training objective $\mathcal{L}_{PIENet}$ for PIENet is a combination of L1, perceptual, adversarial, and codebook losses for generating visually-pleasing and realistic enhanced imaging results:
\begin{multline}
\label{eq:cacloss}
\mathcal{L}_{PIENet} = {\Vert{F_{PIENet}(I_{ab})-I_{gt}}\Vert_1} + \lambda'_{per}\mathcal{L}_{per}\\ 
+ \lambda'_{adv}\mathcal{L}_{adv} + \lambda'_{codebook}\mathcal{L}_{codebook},
\end{multline}
where $F_{PIENet}$ is the PIENet model, and $\{\lambda'_{per}, \lambda'_{adv}, \\ \lambda'_{codebook}\}$ is the other set of hyperparameters for balancing the training process of the image enhancement stage.
The forms of $\mathcal{L}_{per}$, $\mathcal{L}_{adv}$, $\mathcal{L}_{codebook}$ are kept the same as those in pre-training.

\subsection{Training details}
All the training processes of our work are implemented on a single NVIDIA A800 GPU through the Adam optimizer with $\beta_{1} {=} 0.9, \beta_{2} {=} 0.99$, with a batch size of $8$. 
Similar to~\cite{chen2022real}, the codebook is trained via VQ codebook learning with a codebook size of $K{=}1024$, and a dimension of $n{=}512$.
To enable the VQGAN to provide high-quality priors of panoramic images, we additionally train the VQGAN on PALHQ for $200K$ iterations, initialized with the pre-training weights in~\cite{chen2022real}.
With the pre-trained VQGAN frozen, we train the PIENet via $L_{PIENet}$ with a fixed learning rate of $2e{-}4$ for $200K$ iterations.
The loss weights are set as $\lambda_{per}{=}1, \lambda_{adv}{=}0.1, \lambda_{codebook}{=}1$, $\lambda'_{per}{=}1$, $\lambda'_{adv}{=}0.01$, and $\lambda'_{codebook}{=}1$, empirically for a stable training process.  
For data augmentation, random crop, flip, and rotation are applied, where the crop size is $256{\times}256$.

\begin{figure*}[h]
  \centering
  \includegraphics[width=0.35\linewidth]{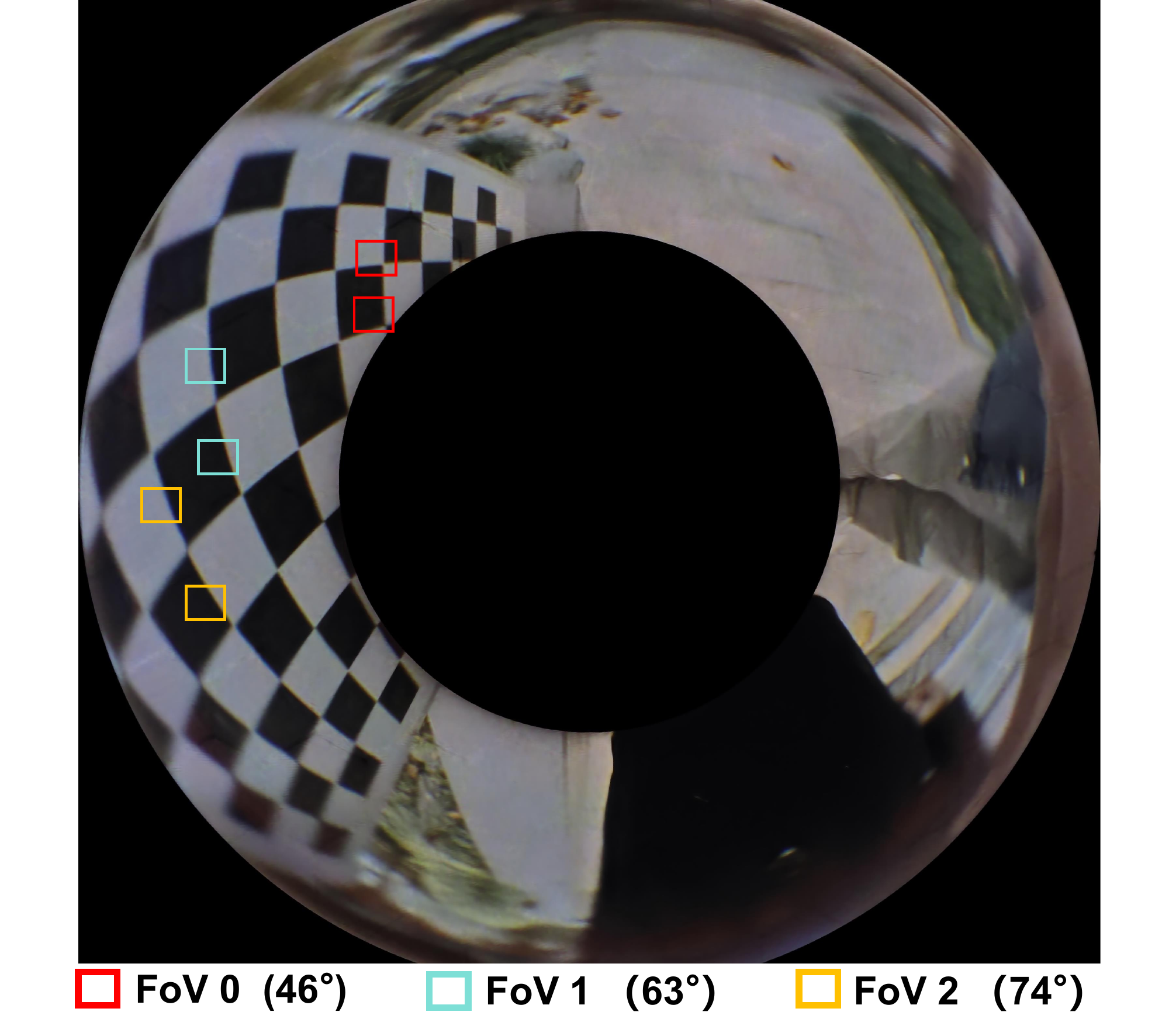}
  \caption{\textbf{Illustration of the knife-edges and specific field-of-view points used for SFR testing.}} 
  \label{fig:mtf_fov}
\end{figure*}

\subsection{Evaluation protocol}
\begin{figure*}[!h]
  \centering
  \includegraphics[width=0.7\linewidth]{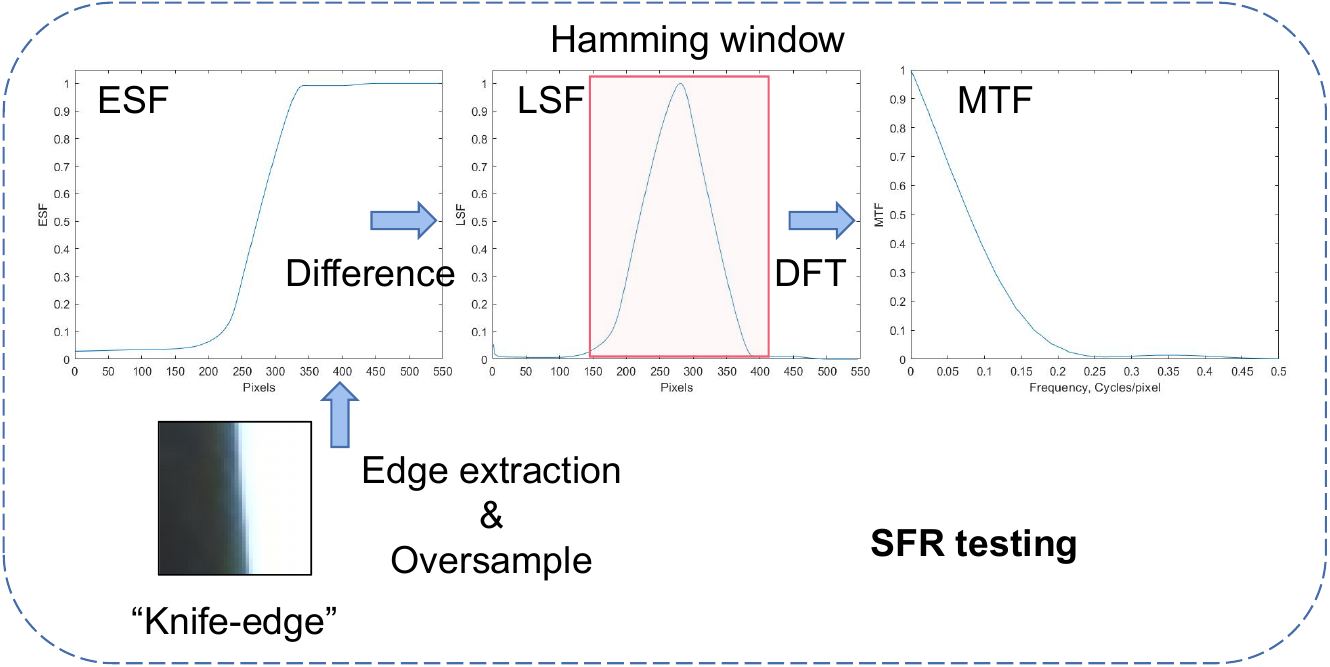}
  \caption{\textbf{A schematic representation of the intermediate processes for SFR testing.} Through SFR testing, we can calculate the MTF of one FoV based on its corresponding ``knife-edge'' image patch.}
  \vskip-2ex
  \label{fig:sfr}
\end{figure*}

For the quality assessment of enhanced MBPIS images in real-world scenarios lacking ground truth references, subject visual quality and modulation transfer function (MTF) derived from checkerboard-pattern-based spatial frequency response (SFR) testing are employed as our evaluation protocol in our main paper.
To conduct SFR testing, we capture real-world images of a checkerboard with MBPIS to obtain ``knife-edge'' images of different FoVs, as shown in Fig.~\ref{fig:mtf_fov}.
For each ``knife-edge'' target within one FoV, the corresponding edge spread function (ESF) can be obtained through oversampling, followed by numerical differentiation and discrete Fourier transform (DFT) to derive the MTF, as mathematically expressed in:
\begin{equation}
\label{eq:mtf}
MTF(\omega) = \mathcal{F}(\frac{d}{dx}ESF(x)), 
\end{equation}
where $\mathcal{F}$ is the DFT process.
The pseudocode delineating the specific workflow of SFR testing is presented in Algorithm~\ref{alg:sfr}, while a schematic representation of the intermediate processes is illustrated in Fig.~\ref{fig:sfr}.

\begin{algorithm}
\caption{The calculation of MTF based on SFR testing.}\label{alg:sfr}
\begin{algorithmic}[1]
\Procedure{CalculateMTF}{$I$}\Comment{Use ``knife-edge'' image patch $I$ to calculate MTF.}
\State $edgeResponse\gets EdgeExtract(I)$\Comment{Extract edge response from the input ``knife-edge''.}
\State $ESF\gets Oversample(edgeResponse)$\Comment{Perform oversampling on the edge response to get edge spread function (ESF).}
\State $LSF\gets Difference(ESF)$ \Comment{Compute the difference of ESF to get line spread function (LSF).}
\State $ windowedLSF\gets ApplyHammingWindow(LSF)$ \Comment{Apply Hamming window to the LSF.}
\State $ MTF\gets DFT(windowedLSF)$ \Comment{Calculate DFT to obtain the MTF.}
\State \textbf{return} $MTF$
\EndProcedure
\end{algorithmic}
\end{algorithm}

\section{System capability for high-quality panoramic imaging}
\subsection{Overall results of enhanced MBPIS image}
\begin{figure*}[!h]
  \centering
  \includegraphics[width=0.7\linewidth]{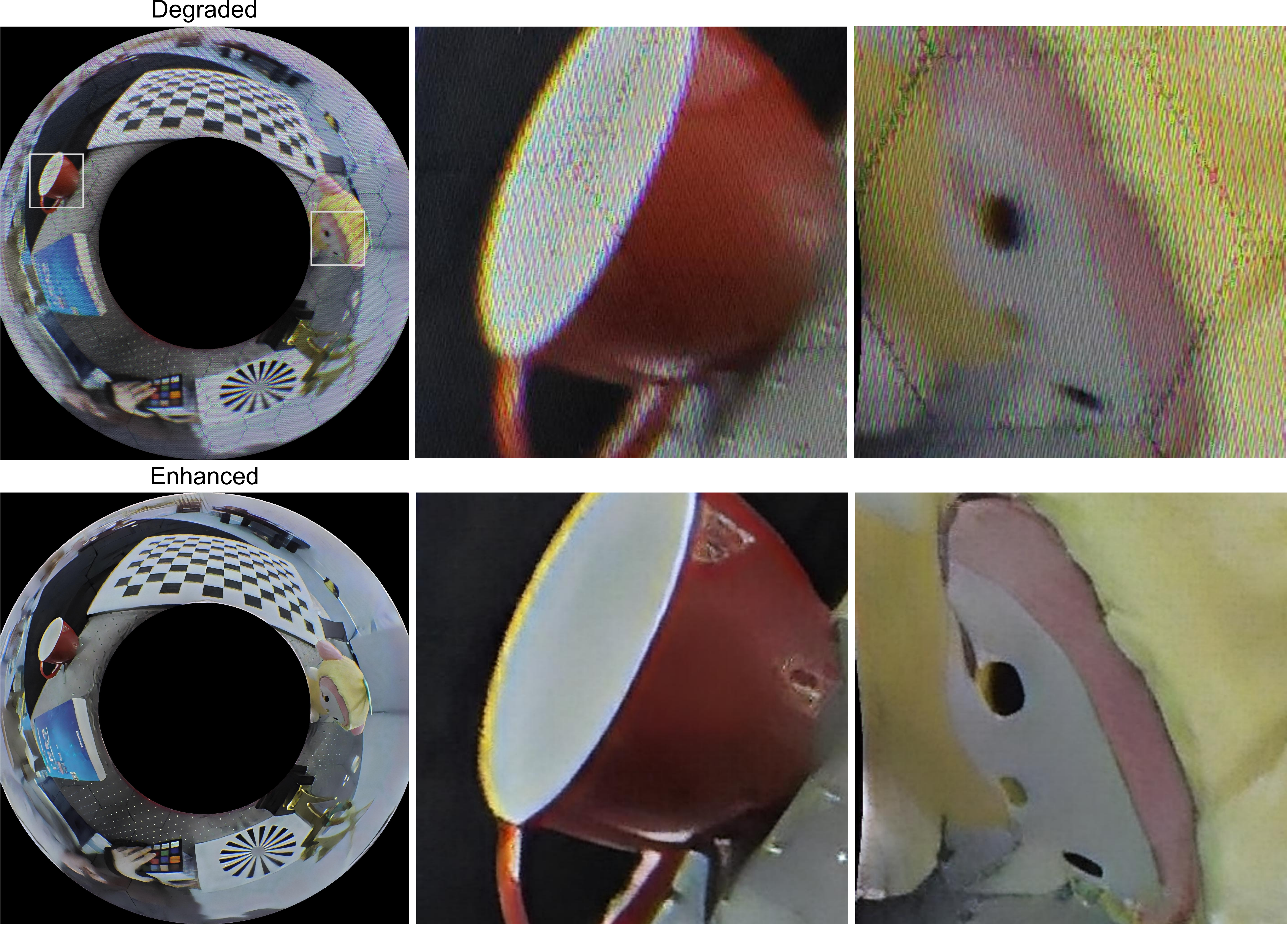}
  \caption{\textbf{Visual comparison between original MBPIS image and enhanced result of PIENet.}}
  \vskip-2ex
  \label{fig:compare_full}
\end{figure*}

Fig.~\ref{fig:compare_full} compares the MBPIS images before and after enhancement by PIENet. The left column shows the full images, while the two right columns present magnified views of selected regions. PIENet effectively suppresses the severe FPN in the original images and mitigates image blur caused by aberrations, resulting in overall sharper and clearer images while recovering most of the details of the original scene.

\subsection{Visual results in multiple scenes}
\begin{figure*}[!h]
  \centering
  \includegraphics[width=0.9\linewidth]{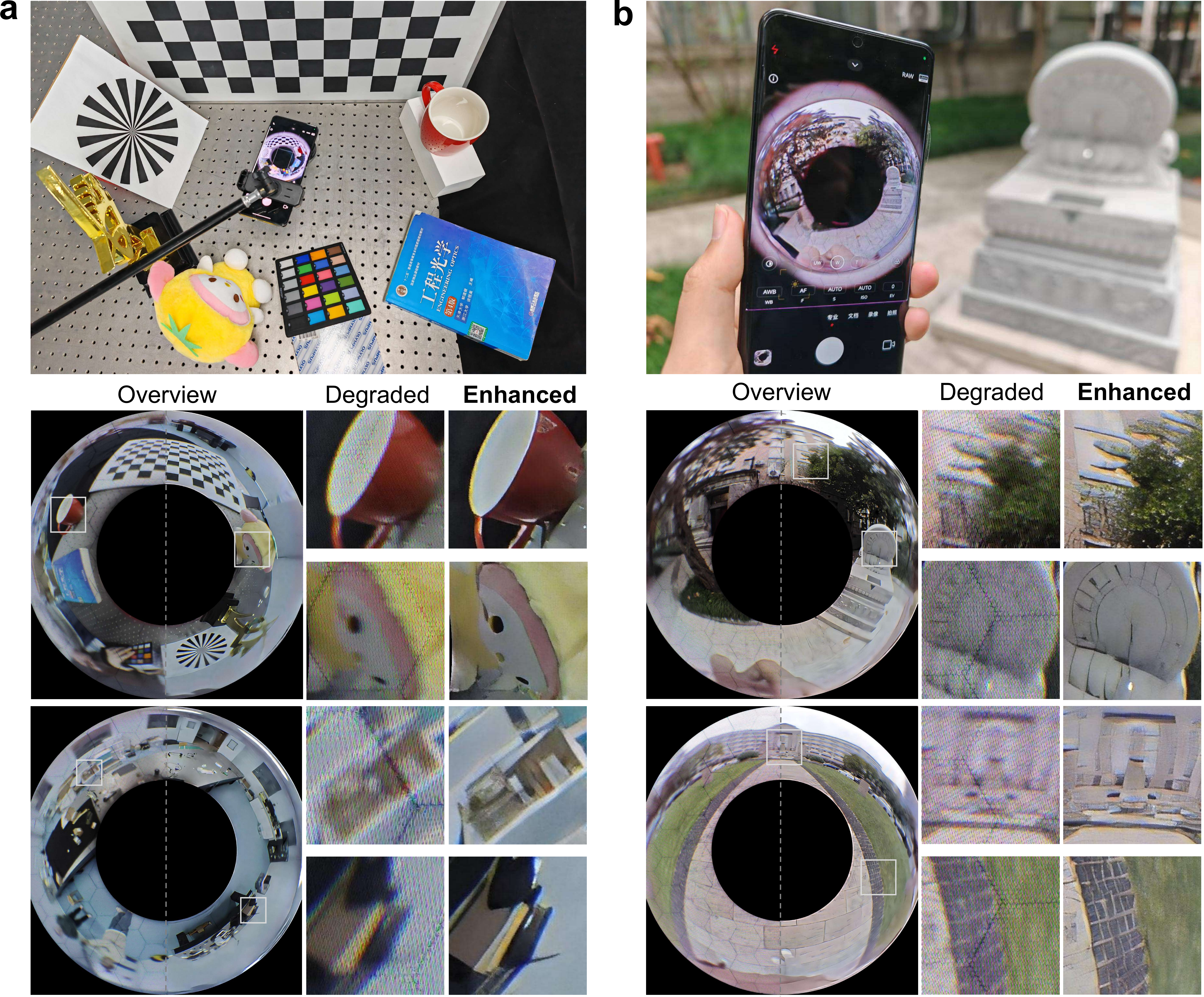}
  \caption{\textbf{Visual results of enhanced MBPIS images in both indoor and outdoor scenes.} At the top, we present the scenarios where MBPIS is utilized for image capture, while below, we display the respective imaging results and image enhancement results for each scene. \textbf{a} Indoor scene. \textbf{b} Outdoor scene.}
  \vskip-2ex
  \label{fig:scene1}
\end{figure*}

Fig.~\ref{fig:scene1} shows the high-quality panoramic imaging results of MBPIS equipped with PIENet for image enhancement in both indoor and outdoor scenes.
Whether in close-up or distant indoor and outdoor scenes, the MBPIS equipped with PIENet consistently delivers high-quality panoramic imaging results: the details across all FoVs are rendered with clarity, free from noise and artifacts, demonstrating the promising application prospects of MBPIS in fields such as panoramic photography and surveillance.

\begin{figure*}[!h]
  \centering
  \includegraphics[width=0.8\linewidth]{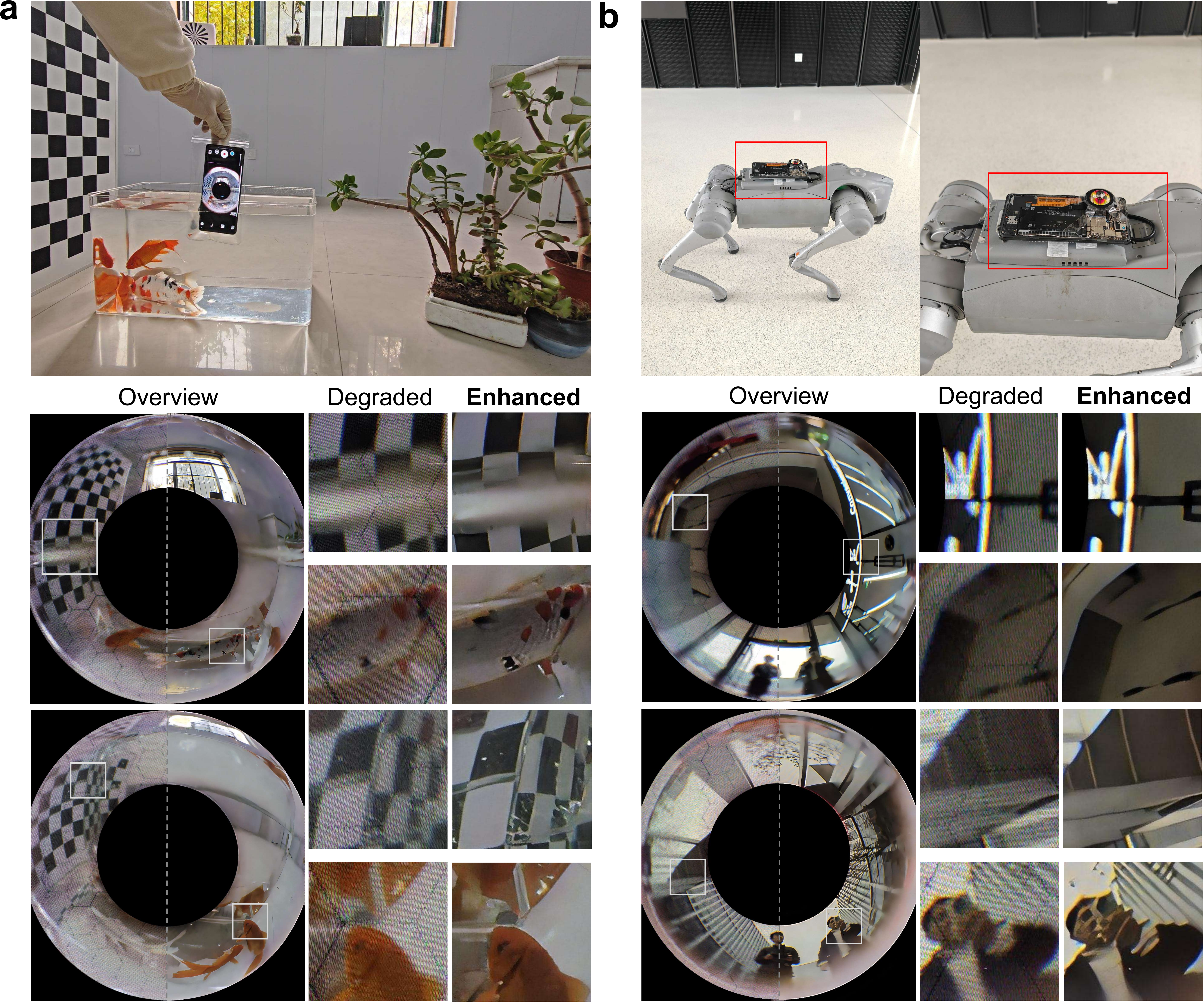}
  \caption{\textbf{Visual results of enhanced MBPIS images in two specific application scenes.} At the top, we present the scenarios where MBPIS is utilized for image capture, while below, we display the respective imaging results and image enhancement results for each scene. \textbf{a} Amphibian visual imaging scene. \textbf{b} Intelligent perception on a quadrupedal robot.}
  \vskip-2ex
  \label{fig:fishdog}
\end{figure*}

Furthermore, as illustrated in Fig.~\ref{fig:fishdog}, leveraging its large FoV, we also demonstrate the application of MBPIS in two additional specialized scenarios: amphibious visual imaging scene and intelligent perception (details in Supplementary Videos 1, 2, and 3). With the image quality enhancement provided by PIENet, MBPIS is capable of delivering high-quality panoramic images for these scenarios, thereby supporting downstream task applications, as detailed in the subsequent section.

\subsection{Ablation study}

\begin{figure*}[!h]
  \centering
  \includegraphics[width=0.7\linewidth]{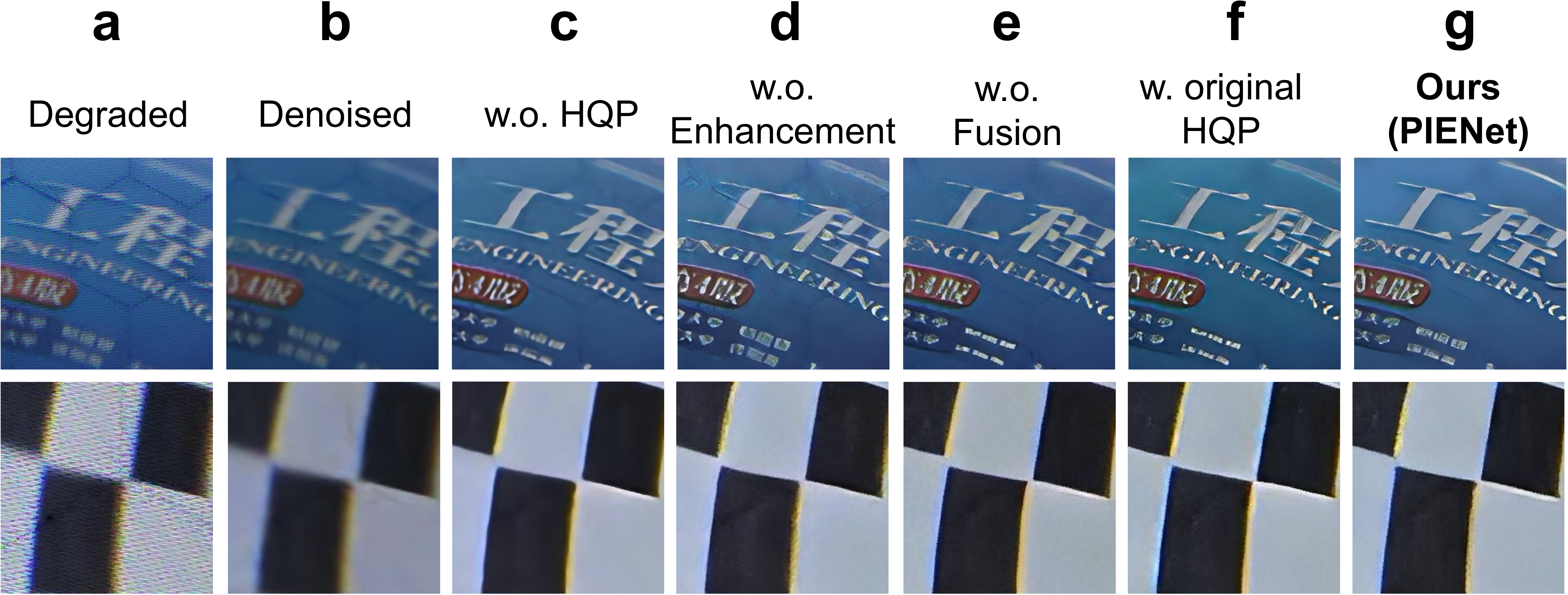}
  \caption{\textbf{Visual results of ablation studies on components of PIENet.} \textbf{a} Degraded MBPIS images. \textbf{b} Denoised MBPIS images to show the optical degradation. We conduct ablations by systematically removing or substituting individual components from the complete PIENet architecture. \textbf{c} Enhanced images produced by PIENet without the guidance of HQP. \textbf{d} Enhanced images produced by PIENet without the RSTB-based enhancement module. \textbf{e} Enhanced images produced by PIENet without the SFT-based fusion module. \textbf{f} Enhanced images produced by PIENet with original HQP pre-trained only on high-quality images of normal FoVs. \textbf{g} Enhanced images produced by our complete PIENet.}
  \vskip-2ex
  \label{fig:ablation}
\end{figure*}

Fig.~\ref{fig:ablation} illustrates the results of our ablation studies on the various components of the proposed PIENet. 
Given that only the results on real-world data hold reference value in our task, we directly evaluate the effectiveness of each experimental setup through visual results.
By comparing \textbf{c-g} with \textbf{a-b}, the effectiveness of our proposed simulation pipeline is initially validated, as all models can successfully address the fixed-pattern noise and significantly mitigate optical degradation, resulting in enhanced image quality with clear details.
Subsequently, the comparison between \textbf{c} and \textbf{g} demonstrates that the guidance from HQP enables the enhanced images to exhibit more realistic textures and details. Furthermore, the comparison between \textbf{f} and \textbf{g} underscores the importance of incorporating high-quality panoramic images in training HQP, as this prevents the introduction of generative artifacts caused by disparities in the FoV distribution.
Moreover, the comparisons among \textbf{d}, \textbf{e}, and \textbf{g} respectively illustrate that both the enhancement module and the fusion module are pivotal components that enable HQP to provide effective guidance. Overall, the complete PIENet yields sharper and more artifact-free image enhancement results.

\section{System capability for video object segmentation}

To further evaluate our minimalist biomimetic computational panoramic imaging system on downstream vision tasks, we combine our imaging system with a quadrupedal robot as an autonomous imaging and perception platform as shown in  Fig~\ref{fig:fishdog}b.
The platform is tested in a complicated environment, where the vibration of the platform poses serious challenges to continuous perception.
The evaluation is conducted on the video object segmentation task, which aims to segment two people who are remotely following the autonomous platform throughout the video.

In practice, we record a video and enhance it by implementing PIENet on each frame.
After the enhancement, the video is used for video object segmentation.
In our case, we use SAM2~\cite{ravi2024sam}, a powerful video segmentation tool, to segment the two persons in the video.
SAM2 takes several positive and negative query points in the initial stage and the refinement stage to segment an object through the entire video.
In practice, we mark 5\~8 positive and negative points for each human object.
The first frame is chosen as the initial stage to mark query points, while the 300\textit{th} frame is used as the refinement stage.
After the feature propagation and segmentation, final masks for each frame of each human object are obtained.
We perform the same segmentation process for both the original and enhanced videos by implementing the same query points in frames with the same index in the video.
Results are shown in Fig.~\ref{fig:dog_segmentation}, where the two subfigures illustrate the two different scenarios (details in Supplementary Video 4).
The second row of each subfigure shows the segmentation results of the corresponding image.
The left side shows the unenhanced images, while the right side shows the enhanced images.
We highlight and zoom in on the obvious differences in the middle.

\begin{figure*}[!h]
  \centering
  \includegraphics[width=0.35\linewidth]{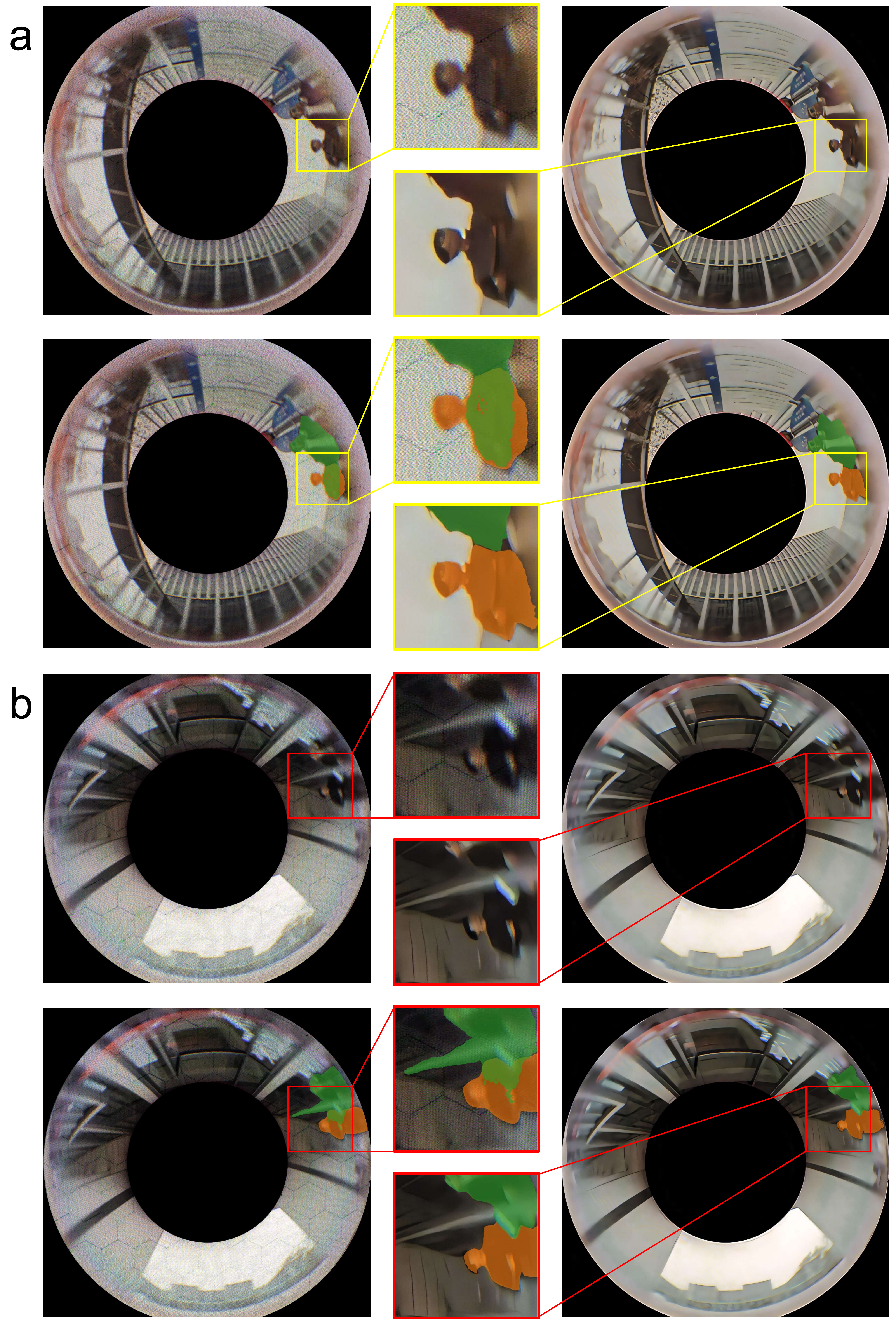}
  \caption{Video object segmentation results on the autonomous platform.}
  \vskip-2ex
  \label{fig:dog_segmentation}
\end{figure*}

Through our proposed image enhancement network PIENet, the optical aberration and the fixed-pattern noise of the original image are well addressed, leading to a clear background and sharp edges.
Building upon the satisfying enhancement results, the video object segmentation results on the final images are more stable and correct than those without using our inspiring computational imaging method.
As shown in Fig.~\ref{fig:dog_segmentation} a, the segmentation model could not find a clear boundary in the low-quality image between the two human masks to be segmented, resulting in an overlapping segmentation area, while the segmentation results in the enhanced image show a clear edge between the two human masks.
Furthermore, false segmentation on the background object is found in the results of low-quality images, as shown in Fig.~\ref{fig:dog_segmentation} b, which supposes that a pillar is a part of one person.
In addition, due to the noise in the low-quality image, the segmentation model could not find a well-aligned body boundary, especially for the small body parts like the head.
The head segmentation results in both Fig.~\ref{fig:dog_segmentation} (a) and (b) all conform to this observation.
The comparisons suggest that the optical aberration and fixed-pattern noise serve as serious negative factors for downstream computer vision tasks, such as video object segmentation.
However, our image enhancement network shows its great ability not only to provide high-quality images but also to eliminate negative impacts on downstream tasks.
As a result, the whole minimalist biomimetic computational panoramic imaging system has excellent imaging effects and can serve downstream tasks well, with a minimalist optical design that only utilizes a single lens to accomplish the panoramic imaging task.

To quantitatively assess the segmentation performance in downstream tasks, we evaluate the masks produced by SAM2 using three metrics:
1) \textit{Predicted Mask Quality}: SAM2's predicted mask IoU, which reflects the model's confidence in its own segmentation. A higher score indicates a higher predicted IoU with the ground-truth mask.
2) \textit{Threshold Stability}: the IoU between masks obtained using different binarization thresholds. This metric measures the sensitivity of the segmentation to threshold changes; a higher score indicates that the predicted mask varies less as the threshold changes.
3) \textit{Prompt Consistency}: the mean IoU between the baseline final mask and four masks generated by shifting all query points left, right, up, and down while keeping the box fixed. Each shift is set to 1\% of the shorter side of the box, with a minimum displacement of one input pixel.

For each metric, we also report the \textbf{Win Rate}, defined as the percentage of tested frames for which the enhanced image achieves a better score than the corresponding degraded image. The results are presented in Table~\ref{tab:SAM_results}.

\renewcommand{\arraystretch}{1.5}
\begin{table*}[h]
    \centering
    \fontsize{7}{7}\selectfont
    \begin{threeparttable}
        \caption{Quantitative comparison of SAM2 segmentation results on degraded and enhanced frames.}
        \label{tab:SAM_results}
        \begin{tabular}{lccc}
            \toprule
             & \textbf{Mask Quality} & \textbf{Threshold Stability} & \textbf{Prompt Consistency} \\
            \midrule
            Degraded & 80.07 & 48.03 & 97.70 \\
            Enhanced & \textbf{90.81} & \textbf{81.68} & \textbf{99.30} \\
            \midrule
            Win Rate (\%) & 99.25 & 99.57 & 95.16 \\
            \bottomrule
        \end{tabular}
    \end{threeparttable}
\end{table*}

As shown in Table~\ref{tab:SAM_results}, compared with the original degraded images, images processed by our restoration network yield higher mask quality, greater threshold stability, and more consistent segmentation responses to prompt perturbations when evaluated with SAM2. The enhanced images achieve better scores in more than 95\% of the tested frames for all three metrics, indicating consistent improvements across the evaluation set.

\end{document}